%% file: wigner_spin_orbit.tex
\documentclass[preprint]{revtex4}
\usepackage[american]{babel}
\usepackage{bm}
\usepackage{amsfonts}
\usepackage{graphicx}
\usepackage{amsmath}
\allowdisplaybreaks[1]

\begin{document}

\title{Phase space methods for the spin dynamics in condensed matter systems}
\author{J\'{e}r\^{o}me Hurst, Paul-Antoine Hervieux, Giovanni Manfredi}
\affiliation{Institut de Physique et Chimie des Mat\'{e}riaux de
Strasbourg, \\ CNRS and Universit\'{e} de Strasbourg\\ BP 43 -- F-67034 Strasbourg Cedex 2, France}

\date{\today}

\begin{abstract}
Using the phase-space formulation of quantum mechanics, we derive a four-component Wigner equation for a system composed of spin-1/2 fermions (typically, electrons) including the Zeeman effect and the spin-orbit coupling. This Wigner equation is coupled to the appropriate Maxwell equations to form a self-consistent mean-field model.
A set of semiclassical Vlasov equations with spin effects is obtained by expanding the full quantum model to first order in the Planck constant. The corresponding hydrodynamic equations are derived by taking velocity moments of the phase-space distribution function. A simple closure relation is proposed to obtain a closed set of hydrodynamic equations.

\end{abstract}
\maketitle

\section{Introduction}

The formulation of quantum mechanics in the phase space was first introduced by Eugene Wigner in 1932 to study quantum corrections to classical statistical mechanics \cite{Wigner}.
The goal was to link the wave function that appears in the Schr\"{o}dinger equation to a pseudo-probability distribution function defined in the classical phase space. This pseudo-probability distribution changes in time according to an evolution equation (Wigner equation) which is somewhat similar to the classical Liouville equation.
Mathematically speaking, the Wigner formulation is based on the Weyl transformation \cite{wigner new,leon cohen book}, which is a general method to transform operators defined in the Hilbert space into phase-space functions.

As it is based on the classical phase space, the Wigner formulation is often a more intuitive approach than the standard Schr\"odinger equation, especially for problems where semiclassical considerations are important. For these reasons, it
is used in many areas of quantum physics, including quantum optics \cite{optic wigner}, semiclassical analysis \cite{sc1,sc2}, electronic transport \cite{elec transport}, nonlinear electron dynamics \cite{jasiak}, and quantum plasma theory \cite{fernando book}. It is also the starting point to construct quantum hydrodynamic equations, which are approximate models obtained by taking velocity moments of the Wigner function. Such models were used in the past to study the electron dynamics in molecular systems \cite{qhy_mol}, metal clusters and nanoparticles \cite{qhy_clust1,qhy_clust2,giovani_paper}, thin metal films \cite{qhy_thin_met}, quantum plasmas \cite{qhy_plasma1,qhy_plasma2}, and semiconductors \cite{qhy_semi_cond}.

The works cited above only considered the charge dynamics and disregarded the spin degrees of freedom. However, it is well known that spin effects (particularly the Zeeman splitting and spin-orbit coupling) can play a decisive role in nanometric systems such as semiconductor quantum dots \cite{puente,serra} and diluted magnetic semiconductors \cite{morandi_NJP_09,morandi_dms_prB}. The coupling between the spin degrees of freedom and the electron orbital motion is of the utmost importance in many experimental studies involving magnetized nano-objects. A particularly interesting example is the ultrafast demagnetization induced by a femtosecond laser pulse in ferromagnetic thin films \cite{Bigot_natphys} -- an effect that is not yet completely elucidated from the theoretical viewpoint.
Recent time-dependant density functional theory (TDDFT) simulations suggest that the spin-orbit coupling plays a central role in the demagnetization process \cite{gross}.

A few theoretical models that include the spin in the Wigner formalism were developed in recent years. One approach \cite{zamanian_NJP10} consists in defining a scalar probability distribution that evolves in an extended phase space, where the spin is treated as a classical two-component variable (related to the two angles on a unit-radius sphere) on the same footing as the position or the momentum. This approach was used to derive a Wigner equation that incorporates spin effects through the Zeeman interaction \cite{zamanian_NJP10}. Semiclassical \cite{zamanian_POP10} and hydrodynamic \cite{asenjo_kinetic_so} spin equations were also derived from those models, including other relativistic effects such as the spin-orbit coupling, the Darwin term, and the relativistic mass correction.

An alternative approach keeps the $2 \times 2$ matrix character of the distribution function \cite{arnold}, so that the orbital and spin dynamics are represented by different Wigner functions. Using this approach, the corresponding Wigner equations were derived from the full Dirac theory \cite{dirac wigner}; however, their complexity makes them unsuitable for applications to condensed matter and nanophysics. A more tractable Wigner equation was derived from the Pauli (instead of Dirac) theory, but only included the Zeeman effect \cite{hurst}.

Both approaches (extended phase space and matrix Wigner function) are equivalent from the mathematical point of view. However, the extended phase-space approach leads to cumbersome hydrodynamic equations that are in practice very hard to solve, either analytically  or numerically, even in the non-relativistic limit. The matrix technique, which is the one used here, separates clearly the orbital motion from the spin dynamics and leads to a simpler and more transparent hydrodynamic model.

In the present paper, we go beyond our previous work \cite{hurst} by including both the Zeeman effect and the spin-orbit coupling, the latter being a relativistic effect to second order in $1/c$. In terms of a semiclassical expansion, these terms are respectively first and second order in $\hbar$. Other relativistic corrections such as the Darwin term or the mass correction are neglected here, although they could be included with relative ease in our model.

First, we will use a gauge invariant formulation of the Weyl transformation \cite{serimaa} and the Moyal product \cite{moyal} to derive a set of Wigner equations describing a system of spin-1/2 fermions.
A self-consistent mean-field model can further be obtained by coupling these Wigner equations to the set of Maxwell equations for the electromagnetic fields, where the sources (charge and current densities) are related to velocity moments of the Wigner function.
A related mean-field model was obtained recently by Dixit et al. \cite{anant}
in the framework of the Schr\"{o}dinger-Pauli equation.

Subsequently, we will derive the corresponding semiclassical limit and obtain the Vlasov equations describing the evolution of an electron gas with spin and semirelativistic effects. In this model, the orbital dynamics is treated classically, whereas the spin is represented as a fully quantum variable.
Finally, the Vlasov equations will be used to derive a hierarchy of hydrodynamic equations by taking velocity moments of the probability distribution function. This is an infinite hierarchy that needs to be closed using some additional physical hypotheses. Although this is relatively easy for spinless systems (where the closure can be obtained by a assuming a suitable equation of state), things are subtler when the spin degrees of freedom are included. Here, we shall use a intuitive closure to obtain a closed set of fluid equations with spin effects.

\section{Quantum mechanics in the phase space}

The basic idea of the phase-space formulation of quantum mechanics is to associate at each operator $\mathcal{\widehat{O}}\left(\bm{\widehat{R}},\bm{\widehat{P}}\right)$, depending on the position and momentum operators $\widehat{R}$ and $\widehat{P}$, a function $\mathcal{O}(\bm{r},\bm{p})$ of the classical phase-space variables $\bm{r}$ and $\bm{p}$. This correspondence is provided by the Weyl transformation \cite{weyl original,leon cohen book}, and is given by:
\begin{align}
\mathcal{\widehat{O}}\left(\bm{\widehat{R}},\bm{\widehat{P}}\right) \equiv &
\int d\bm{r}~d\bm{p} ~\mathcal{O}\left(\bm{r},\bm{p}\right) \mathcal{\widehat{F}}\left(\bm{r},\bm{p}\right),
\label{weyl transform}
\end{align}
where $ \mathcal{\widehat{F}}\left(\bm{r},\bm{p}\right)$ is the Wigner operator defined as
\begin{align}
 \mathcal{\widehat{F}}\left(\bm{r},\bm{p}\right) \equiv & \frac{1}{\left(2\pi \hbar\right)^{6}} \int d\bm{u} ~d\bm{v} \exp \left[ \frac{i}{\hbar} \left( \bm{u} \cdot \left( \bm{\widehat{P}} - \bm{p} \right) +  \bm{v} \cdot \left( \bm{\widehat{R}} - \bm{r} \right)\right) \right].
\label{Wigner operator}
\end{align}
The inverse of the Weyl transformation can be deduced \footnote{For the demonstration, we use the following properties of the Wigner operator: $\textrm{tr} \left[  \widehat{\mathcal{F}}\left(\bm{r},\bm{p}\right) \widehat{\mathcal{W}}\left(\bm{r'},\bm{p'}\right) \right] = \frac{1}{\left(2 \pi \hbar \right)^{3}} \delta \left( \bm{p} - \bm{p'} \right) \delta \left( \bm{r} - \bm{r'} \right).$} from the above definition:
\begin{align}
\mathcal{O}\left(\bm{r},\bm{p}\right) = \left(2 \pi \hbar \right)^{3} \textrm{tr} \left[  \mathcal{\widehat{F}}\left(\bm{r},\bm{p}\right) \mathcal{\widehat{O}}\left(\bm{\widehat{R}},\bm{\widehat{P}}\right) \right],
\label{inverse weyl transform}
\end{align}
where $\textrm{tr}$ denotes the trace.

Considering a system in a statistical distribution described by the density operator $\widehat{\rho}$, Eq. \eqref{weyl transform} can be used to determine the mean value of an arbitrary operator:
\begin{align}
\nonumber \langle \mathcal{\widehat{O}}\left(\bm{\widehat{R}},\bm{\widehat{P}}\right)  \rangle =&
~\textrm{tr} \left[ \mathcal{\widehat{O}}\left(\bm{\widehat{R}},\bm{\widehat{P}}\right) \widehat{\rho} \right] = \int d\bm{r} ~ d\bm{p}~ \mathcal{O}\left(\bm{r},\bm{p}\right) \textrm{tr} \left[\mathcal{\widehat{F}}\left(\bm{r},\bm{p}\right) \widehat{\rho} \right].
\end{align}
The Wigner function is then defined as the phase-space function associated to the density operator
\begin{align}
f\left(\bm{r},\bm{p}\right) = \textrm{tr} \left[\mathcal{\widehat{F}}\left(\bm{r},\bm{p}\right) \widehat{\rho} \right] =\frac{1}{\left( 2 \pi \hbar \right)^{3}} \int d\bm{\lambda} \exp \left( \frac{i}{\hbar} \bm{\lambda} \cdot \bm{p} \right) \left\langle \bm{r} - \frac{1}{2}\bm{\lambda} | \widehat{\rho} | \bm{r} + \frac{1}{2}\bm{\lambda} \right\rangle.
\label{wigner function density operator}
\end{align}
The Wigner function obeys to the following equation of motion (Wigner equation):
\begin{align}
 i \hbar \frac{\partial f }{\partial t} =  \left\{  \mathcal{H} ,  f \right\}_{\star},
\label{evoultion equation wigner moyal braket}
\end{align}
where the last term is referred to as the Moyal bracket
\begin{align}
\left\{ A(\bm{r},\bm{p}) , B(\bm{r},\bm{p}) \right\}_{\star} = 2i \sin \left[\frac{\hbar}{2} \left( {}^L \! \partial_{i} {}^R \! \partial_{p_{i}} - {}^L \! \partial_{p_{j}} {}^R \! \partial_{j} \right) \right] \left(A(\bm{r},\bm{p}), B(\bm{r},\bm{p}) \right).
\label{Moyal bracket}
\end{align}
The indexes $L$ and $R$ mean that the derivative acts only on the left or on the right term in the parenthesis.
The Wigner equation is the analogue of the density matrix evolution equation in the operator representation of quantum mechanics: $i\hbar \partial_t \widehat{\rho} = \left[ \widehat{H},\widehat{\rho} \right]$, sometimes called the Von Neumann equation.
The Moyal brackets can be easily developed as a power series of $\hbar$, which makes the Wigner formulation particularly interesting to study the semiclassical limit. The lowest order term leads to the standard Poisson brackets and to the equations of classical mechanics.

In Eq. \eqref{evoultion equation wigner moyal braket}, $\mathcal{H}$ is the phase space function associated to the Hamiltonian operator $\widehat{\mathcal{H}}$ of the system, and they are related by Eq. \eqref{weyl transform}. In order to determine the phase space function of any arbitrary operator $\mathcal{\widehat{O}}\left(\bm{\widehat{R}},\bm{\widehat{P}}\right)$, one should apply the Weyl correspondence rule \citep{serimaa,leon cohen book}, defined as follows: (i) first symmetrize the operator $\mathcal{\widehat{O}}\left(\bm{\widehat{R}},\bm{\widehat{P}}\right)$ with respect to the position and the momentum operators $\bm{\widehat{R}}$ and $\bm{\widehat{P}}$; (ii) then replace $\mathcal{\widehat{O}}\left(\bm{\widehat{R}},\bm{\widehat{P}}\right)$ by their associated classical variables. For instance, for the operator $\widehat{P}_{x}\widehat{X}$ one finds
\begin{align}
 \widehat{P}_{x} \widehat{X} = \frac{1}{2}\left(  \widehat{P}_{x} \widehat{X} + \widehat{X} \widehat{P}_{x} \right)   - \frac{i\hbar}{2} \rightarrow  xp_{x} - \frac{i\hbar}{2},
\end{align}
where use has been made of the commutator $[\widehat{P}_{x},\widehat{X}]=\hbar/i$.
We note that the Weyl correspondence defined above is not unique, and one could have defined other rules leading to different phase-space function, such as the the Husimi representation \cite{husimi}.

In the case of a spinless particle moving in a scalar potential $V(\bm{r})$, the Wigner evolution equation \eqref{evoultion equation wigner moyal braket} reads as
\begin{align}
\frac{\partial f\left(\bm{r},\bm{p},t\right)}{\partial t} + \frac{1}{m}\bm{p}\cdot \bm{\nabla}f\left(\bm{r},\bm{p},t\right)
&=
-\frac{i}{\hbar}\frac{1}{\left( 2 \pi \hbar \right)^{3}} \int d\bm{\lambda} d\bm{p'} \exp \left[\frac{ i \left(\bm{p}- \bm{p'} \right) \cdot \bm{\lambda}}{\hbar} \right]\nonumber \\
&~~~
\times \left[ V\left( \bm{r} + \frac{\bm{\lambda}}{2} \right) - V\left( \bm{r} -\frac{ \bm{\lambda}}{2} \right) \right] f \left(\bm{r},\bm{p'},t\right).
\label{wigner equation sans spin sans B}
\end{align}
However, complications arise when  we want to include magnetic interactions. It is well known that in presence of magnetic fields one should use the kinetic momentum operator $\widehat{\bm{\Pi}} = \widehat{\bm{P}} -q \widehat{\bm{A}}$ instead of $\widehat{\bm{P}}$. This situation cannot be addressed by simply replacing $\widehat{\bm{P}}$ with $\widehat{\bm{\Pi}}$ in the Weyl transformation. Indeed it can be easily proven that with such substitution the Wigner function, Eq. \eqref{wigner function density operator}, is not gauge invariant. As spin effects, such as the Zemann interaction or the spin-orbit coupling, strongly depend on the magnetic field, it is of paramount importance to work with a gauge invariant formulation of the Weyl transformation.
A gauge independent definition of the Wigner function was first introduced by Stratonovich \cite{stratonovich}:
\begin{align}
f\left(\bm{r},\bm{v}, t \right) &=
\left(\frac{ m}{2 \pi \hbar}\right)^{3} \int d\bm{\lambda} \exp \left[ \frac{i \bm{\lambda}}{\hbar} \cdot \left(m \bm{v} + q \int_{-1/2}^{1/2} d\tau \bm{A}\left(\bm{r} + \tau \bm{\lambda} \right) \right) \right] \left\langle \bm{r} - \frac{\bm{\lambda}}{2} \vline\, \widehat{\rho} \,\vline\, \bm{r} + \frac{\bm{\lambda}}{2} \right\rangle.
\label{one body dis function exp with magnetic field}
\end{align}
where the momentum $\bm{p}$ was replaced by $m \bm{v} + q \int_{-1/2}^{1/2} d\tau \bm{A}\left(\bm{r} + \tau \bm{\lambda}\right)$.

To be consistent with this new definition of the Wigner function, one should also modify the Weyl correspondence rule \cite{serimaa}. The procedure is identical, except that one should use $\widehat{\bm{\Pi}}$ instead of $\widehat{\bm{P}}$. The main difference is that one must also symmetrize operators with respect to the different component of $\widehat{\bm{\Pi}}$, because they do not commute, i.e., $\left[\widehat{\Pi}_{i},\widehat{\Pi}_{j}\right] = i\hbar q \epsilon_{ijk} B_{k} ( \widehat{\bm{R}})$, where $\epsilon_{ijk}$ is the Levi-Civita symbol. The classical phase space variable associated to the kinetic momentum operator is the linear momentum $\widehat{\bm{\Pi}} \rightarrow \bm{\pi} = m\bm{v}$.\\
The Moyal product defined in Eq. \eqref{Moyal bracket} is also modified in the presence of magnetic fields. A gauge invariant Moyal product was derived by M\"{u}ller \cite{muller}, and reads
\begin{align}
A(\bm{x},\bm{\pi}) \star C(\bm{r},\bm{\pi}) &= \exp \left[ i \hbar \mathcal{L} + ie \sum_{n=1}^{\infty} \hbar^{n} \mathcal{L}_{n} \right] \left( A(\bm{r},\bm{\pi}) , C(\bm{r},\bm{\pi})\right),
\label{definition moyal product with magnetic field}
\end{align}
where $\mathcal{L}$ is the operator corresponding to the free magnetic field case (here and in the following, we use Einstein's summation convention) :
\begin{align}
 \mathcal{L} \left( A(\bm{r},\bm{\pi}), C(\bm{r},\bm{\pi}) \right) &= \frac{1}{2} \left(   {}^L \! \partial_{i} {}^R \! \partial_{\pi_{i}} - {}^R \! \partial_{j} {}^L \! \partial_{\pi_{j}}   \right)  \left( A(\bm{r},\bm{\pi}), C(\bm{r},\bm{\pi}) \right)
 \label{operateur L}
\end{align}
and $\mathcal{L}_{n}$ is a new operator that depends on the magnetic field:
\begin{align}
 \mathcal{L}_{n}\left( A(\bm{r},\bm{\pi}), C(\bm{r},\bm{\pi})\right) &= \left(\frac{i}{2}\right)^{n+1} \frac{\epsilon_{jlr} }{\left(n+1\right)^{2} n!} \left(\partial ^{n-1}_{i_1 ... i_{n-1}} B_{r} \right)    {}^L \! \partial_{\pi_{j}} {}^R \! \partial_{\pi_{l}} \sum_{p=1}^{n}
 \begin{pmatrix}
 n+1 \\ p
 \end{pmatrix}
 g(n,p)
\nonumber \\
&~  {}^L \! \partial_{\pi_{i_1}} \cdots   {}^L \! \partial_{\pi_{i_{p-1} }} {}^R \! \partial_{\pi_{i_p}} \cdots  {}^R \! \partial_{\pi_{i_{n-1} }}  \left( A(\bm{r},\bm{\pi}), C(\bm{r},\bm{\pi}) \right).
\label{operateur ln}
\end{align}
with $g(n,p) = \left[\left(1-\left(-1\right)^{p}\right) \left(n+1\right) - \left(1-\left(-1\right)^{n+1}\right) p \right]$.
This new definition of the Moyal product makes the calculation of the evolution equation much more cumbersome than in the unmagnetized case. Its great advantage is that it ensures that the final equations of motion are gauge invariant.

\section{Derivation of the spin Wigner model}

We consider an ensemble of fermions in the presence of an electromagnetic field $\bm{E}$, $\bm{B}$. We denote the  Schr\"odinger wave function of the $\mu-$th particle state by
\begin{equation}
\Psi_{\mu}(\bm{r},t) =\Psi_{\mu}^{\uparrow}(\bm{r},t)\, \left|\uparrow \right\rangle +\Psi_{\mu}^{\downarrow}(\bm{r},t)\, \left|\downarrow \right\rangle, \label{pauli-spinor}
\end{equation}
where $ \Psi_{\mu}^{\uparrow}(\bm{r},t)$  and $\Psi_{\mu}^{\downarrow}(\bm{r},t)$ are respectively the spin-up and spin-down components of the wave function. The  evolution of the system is governed by the Pauli-Schr\"odinger equation
\begin{align}
& i \hbar \frac{\partial \Psi_{\mu} \left( \bm{r},t\right) }{\partial t} = \mathcal{H}\Psi_{\mu}  \left( \bm{r},t\right),\\
&\mathcal{H} = \left(\frac{\widehat{\bm{\Pi}}^{2}}{2m} + V \right) \sigma_{0}  + \left[ \mu_{B}  \widehat{\bm{B}} + \frac{\mu_{B}}{4mc^{2}} \left( \widehat{\bm{E}} \times \widehat{\bm{\Pi}}   -  \widehat{\bm{\Pi}} \times \widehat{\bm{E}} \right) \right] \cdot \bm{\sigma}.
\label{Pauli equation}
\end{align}
Here, $\mu_{B}=e\hbar/2m$ is the Bohr magneton, $ \bm{ \sigma} $ is the vector made of the $2 \times 2$ Pauli matrices, and $\sigma_{0}$ is the $2 \times 2$ identity matrix. $V$ and $\bm{A}$ are, respectively, the scalar and vector potential.
Equation \eqref{Pauli equation} can be derived from the Dirac equation by means of a Foldy-Wouthuysen  transformation \cite{hinsch,strange}. This semirelativistic  development leads to plenty of terms that couple the spin to the charge dynamics. In this work we only keep terms up to second order in $1/c$, where $c$ is the speed of light in vacuum, namely the Zeeman interaction (order $1/c^0$) and the spin-orbit coupling (order $1/c^2$). We neglect however the Darwin term and the relativistic mass correction, which are also second order effects.

Without spin, the Wigner function is a scalar function related to the density matrix $\rho$ through Eq. \eqref{one body dis function exp with magnetic field}. This definition can be generalized as follows to take into account the spin degrees of freedom:
\begin{align}
F\left(\bm{r},\bm{v}, t \right) &=
\left(\frac{ 1}{2 \pi \hbar}\right)^{3} \int d\bm{\lambda} \exp \left[ \frac{i \bm{\lambda}}{\hbar} \cdot \left(m \bm{v} + q \int_{-1/2}^{1/2} d\tau \bm{A}\left(\bm{r} + \tau \bm{\lambda} \right) \right) \right]
\rho  ( \bm{r} - \bm{\lambda} / 2, \bm{r} + \bm{\lambda} / 2,t),
\label{wignerfunction}
\end{align}
where, for particles with spin 1/2, $F$ is a $2\times 2$ matrix.  The elements of the density matrix $\rho ^{\eta \eta '} (\bm{r},\bm{r}',t) $ where $\eta =  \uparrow , \downarrow  $, are given by
\begin{equation}
\rho ^{\eta \eta '} (\bm{r},\bm{r}') = \sum _{\mu} \Psi_{\mu}^{\eta}(\bm{r},t) \Psi_{\mu}^{\eta ' *}(\bm{r}',t).
\label{matrix density}
\end{equation}
In order to study the macroscopic properties of the system, it is convenient to project $F$ onto the Pauli basis set \cite{barletti_03,morandi_JPA_11}
\begin{equation}
 F= \frac{1}{2}\sigma _0 f_0 + \frac{1}{\hbar}\bm{f} \cdot \bm{\sigma},
\label{change basis wigner function}
\end{equation}
where
\begin{equation}
f _{0} = \textrm{tr} \left\{ F \right\}  = f ^{\uparrow \uparrow} + f ^{\downarrow \downarrow}, ~~~~
\bm{f}   = \frac{\hbar}{2} \textrm{tr} \left( F \bm \sigma \right)
\label{def f0 f_vec}
\end{equation}
and tr denotes the trace.
With this definition, the particle density $n $ and the spin polarization $\bm{S} $ of the electron gas are easily expressed by the moments of the pseudo-distribution functions  $f_0$ and $\bm f$:
\begin{eqnarray}
n(\bm{r},t)
&=&\sum _{\mu}  \left|  \Psi_{\mu}^\dagger (\bm{r},t) \right| ^2
=
\int f_{0} (\bm{r},\bm{v}, t) d\bm{v}, \label{def n} \\
\bm S (\bm{r},t)
&=&
\frac{\hbar}{2} \sum_{\mu}  \Psi_{\mu}^\dagger(\bm{r},t)\bm \sigma  \Psi_{\mu}(\bm{r},t)
=
\int \bm f (\bm{r},\bm{v}, t)  d\bm{v}.\label{def S}
\end{eqnarray}

In this representation, the Wigner functions have a clear physical interpretation: $f_{0}$ is related to the total electron density (in phase space), whereas $f_{k}$ ($k=x,y,z$) is related to the spin polarization in the direction $k$.
In other words, $f_{0}$ represents the probability to find an electron at one point of the phase space at a given time, while $f_{k}$ represents the probability to have a spin-polarization probability in the direction $k$ for this electron.

There exist different ways to include the spin in the Wigner formalism other than the one described above. For instance, Brodin et al. \cite{zamanian_NJP10}, introduced an extended phase space ($\bm{r}, \bm{v}, \bm{s}$) where $\bm{s}$ is a unitary vector that defines the spin direction. The corresponding probability distribution is a scalar function of the extended phase-space variables.
This is in contrast with our approach, where the spin is treated as a fully quantum variable (evolving in a two-dimensional Hilbert space).
Nevertheless, the two approaches are equivalent, as shown in Ref. \cite{hurst}.
The correspondence relations between our distribution functions $f_{0}(\bm{r}, \bm{v}, t)$ and $f_{k}(\bm{r}, \bm{v}, t)$ and the scalar distribution used by Zamanian et al. [53] $f_Z(\bm{r}, \bm{v}, \bm{s}, t)$ read as:
\begin{align}
f_{0} &= \int f_Z d^{2}\bm{s},~~~~ \bm{f} = 3\int \bm{s} f_Z d^{2}\bm{s}.
\end{align}

Let us now turn to the evolution equation for the Wigner functions $f_{0}(\bm{r}, \bm{v}, t)$ and $f_{k}(\bm{r}, \bm{v}, t)$.
After some tedious calculations, developed in the Supplementary Material,  Eq. \eqref{evoultion equation wigner moyal braket} leads to the following Wigner equations:
\begin{align}
&
\frac{\partial f_{0}}{\partial t}
+
\frac{1}{m}\left(\bm{\pi} + \bm{\Delta \widetilde{\pi}}  \right)\cdot \bm{\nabla}f_{0}
-
\frac{e}{m} \left[ m\widetilde{\bm{E}} + \left(\bm{\pi} + \bm{\Delta \widetilde{\pi}}  \right) \times \widetilde{\bm{B}} \right]_{i} \partial_{\pi_{i}} f_{0}  \nonumber \\
&~
- \mu_{b} \bm{\nabla}\left( \widetilde{\bm{B}} -\frac{1}{2mc^{2}} \bm{\pi}  \times \widetilde{\bm{E}} \right)_{i}  \cdot \bm{\nabla_{\pi}} f_{i}
+
\frac{\mu_{b}}{4mc^{2}}\left[\left(\bm{E}_{+} + \bm{E}_{-}  \right)    \times \bm{\nabla} \right]  \cdot \bm{f} \nonumber \\
&~
-
\frac{\mu_{b}e}{2mc^{2}}
\left[\widetilde{\bm{E}}  \times \left[ \widetilde{\bm{B}}  \times \bm{\nabla}_{\pi}\right] \right]\cdot \bm{f}
-
\frac{\mu_{b}}{2mc^{2}}\frac{i}{\hbar}
\left[\bm{\Delta \widetilde{\pi}}  \times   \left(\bm{E}_{+} - \bm{E}_{-}  \right) \right]  \cdot \bm{f}
= 0, \label{wigner equation f0 condensee avec spin avec B}
\end{align}
\begin{align}
&
\frac{\partial f_{k}}{\partial t}
+
\frac{1}{m}\left(\bm{\pi} + \bm{\Delta \widetilde{\pi}}  \right)\cdot \bm{\nabla}f_{k}
-
\frac{e}{m} \left[ m\widetilde{\bm{E}} + \left(\bm{\pi} + \bm{\Delta \widetilde{\pi}}  \right) \times \widetilde{\bm{B}} \right]_{i} \partial_{\pi_{i}} f_{k}  \nonumber \\
&~
- \mu_{b} \bm{\nabla}\left( \widetilde{\bm{B}} -\frac{1}{2mc^{2}} \bm{\pi}  \times \widetilde{\bm{E}} \right)_{k}  \cdot \bm{\nabla_{\pi}} f_{0}
+
\frac{\mu_{b}}{4mc^{2}}\left[\left(\bm{E}_{+} + \bm{E}_{-}  \right)  \times \bm{\nabla} \right]_{k} f_{0} \nonumber \\
&~
-
\frac{\mu_{b}e}{2mc^{2}}
\left[\widetilde{\bm{E}}  \times \left[ \widetilde{\bm{B}}  \times \bm{\nabla}_{\pi}\right] \right]_{k} f_{0}
-
\frac{\mu_{b}}{2mc^{2}}\frac{i}{\hbar}
\left[\bm{\Delta \widetilde{\pi}}  \times   \left(\bm{E}_{+} - \bm{E}_{-}  \right)  \right]_{k} f_{0} \nonumber \\
&~
-\frac{e}{2m} \left[ \left(\bm{B}_{+} + \bm{B}_{-}  - \frac{1}{2mc^{2}} \left(\bm{\pi} + \bm{\Delta \widetilde{\pi}} \right)   \times \left(\bm{E}_{+} + \bm{E}_{-}  \right)\right) \times \bm{f}\right]_{k}\nonumber \\
&~
 +
 \frac{\mu_{b}}{2mc^{2}}\frac{i}{2}
 \left[ \left( \left( \bm{E}_{+} - \bm{E}_{-}  \right) \times \left( \bm{\nabla} -e  \widetilde{\bm{B}} \times \bm{\nabla_{\pi}} \right)  \right) \times \bm{f} \right]_{k} = 0,
\label{wigner equation f condensee avec spin avec B}
\end{align}
where $\bm{\Delta \widetilde{\pi}} $ depends of the magnetic field and corresponds to a quantum shift of the velocity
\begin{align}
\bm{\Delta \widetilde{\pi}} = -i\hbar e \partial_{\bm{\pi}} \times \left[ \int^{1/2}_{-1/2} d\tau \,\tau \bm{B} \left( \bm{r} + i\hbar \tau \partial_{\bm{\pi}} \right) \right]
\label{def shift vitesse quantique}
\end{align}
and  $\widetilde{\bm{E}}$ and $\widetilde{\bm{B}}$  are written in terms of the electric and magnetic fields
\begin{align}
\widetilde{\bm{E}} &= \int^{1/2}_{-1/2} d\tau \bm{E} \left( \bm{r} + i\hbar \tau \partial_{\bm{\pi}} \right),~~~~
\widetilde{\bm{B}} =  \int^{1/2}_{-1/2} d\tau  \bm{B} \left( \bm{r} + i\hbar \tau \partial_{\bm{\pi}} \right). \label{def E B quantique}
\end{align}
The subscripts $\pm$ means that the corresponding quantity is evaluated at a shifted position $ \bm{r} \pm i\hbar \partial_{\bm{\pi}} /2$.
This particularly illuminating form of the Wigner equations was proposed by Serimaa et al. \cite{serimaa} for the case of a charged particle without spin evolving in an external electromagnetic field.

Equations \eqref{wigner equation f0 condensee avec spin avec B}-\eqref{wigner equation f condensee avec spin avec B} can be viewed as a generalization of those obtained in our previous work \cite{hurst}, where only the Zeeman interaction was included. The latter has two effects: the first is to couple the spin to the orbital dynamics through the gradient of the magnetic field (terms $\mu_{b} \bm{\nabla} \widetilde{B}_{k} \cdot \bm{\nabla_{\pi}}$ in the equations); the second effect is the spin precession around an effective magnetic field [terms ($\bm{B}_{+} + \bm{B}_{-}) \times \bm{f}]$.
In addition, many new terms appear due to the spin-orbit interaction, which can be easily identified because they are proportional to $1/c^{2}$. Some of these terms couple the spin to the orbital dynamics, while others provide corrections to the spin precession or the Lorentz force. The physical origin of all these terms will appear clearly in the next session, when we discuss the semiclassical limit of the Wigner equations.

Equations \eqref{wigner equation f0 condensee avec spin avec B}-\eqref{wigner equation f condensee avec spin avec B} can be used, in the context of a mean-field approach, to describe the self-consistent spin dynamics of an ensemble of interacting electrons. In this case, the electric and the magnetic fields are solutions of the Maxwell equations:
\begin{equation}
\begin{array}{lcl}
\bm{\nabla} \cdot \bm{E} &=& \frac{\rho}{\epsilon_{0}} - \frac{\bm{\nabla} \cdot \bm{P} }{\epsilon_{0}},  \\
\bm{\nabla} \cdot \bm{B} &=& 0, \\
\bm{\nabla} \times \bm{E} &= &-\frac{\partial \bm{B}}{\partial t},  \\
\bm{\nabla} \times \bm{B} &=& \mu_{0} \bm{j} + \mu_{0} \epsilon_{0}\frac{\partial \bm{E}}{\partial t}  + \mu_{0} \frac{\partial \bm{P}}{\partial t} + \mu_{0}\bm{\nabla} \times \bm{M},
\end{array}
\label{maxwell equations}
\end{equation}
where we introduced some relativistic corrections to the source terms, namely a spin magnetization $\bm{M}$, a spin polarization $\bm{P}$ and a new contribution to the current density, see Eq. \eqref{def current generalized}.
These corrections appear when one considers an expansion in $1/c$ of the Dirac-Maxwell equations and are consistent with the Hamiltonian of Eq. \eqref{Pauli equation}, as was shown recently using a Lagrangian method \cite{anant,manfredi_maxwell}.
Using Eq. \eqref{one body dis function exp with magnetic field},  We can transpose their results to our formulation, which yields
\begin{align}
\rho   &= -e \int f_{0} d\bm{v},\label{def density}  \\
\bm{j} &= -e \left[ \int  \bm{v} f_{0} d\bm{v} + \frac{\bm{E} \times \bm{M} }{2mc^{2}}   \right], \label{def current generalized} \\
\bm{M} &= - \mu_{B}\int \bm{f} d\bm{v}, \label{def magnetization}  \\
\bm{P} &=  -\frac{\mu_{B}}{2c^{2}} \int \bm{v} \times \bm{f} d\bm{v} .
\label{def spin polarization}
\end{align}
This mean-field approach could in principle be extended, in the spirit of density functional theory, to include exchange and correlation effects by adding suitable potentials and fields that are functionals of the electron density  \cite{metal_films_2}.

\section{Semiclassical limit and spin Vlasov model}

The form of Eqs. \eqref{wigner equation f0 condensee avec spin avec B}-\eqref{wigner equation f condensee avec spin avec B} is particularly useful to study the semiclassical limit of the model. Indeed, we can easily expand  $\widetilde{\bm{E}}$, $\widetilde{\bm{B}} $ and $\bm{\Delta \widetilde{\pi}} $  as a power series of $\hbar$
\begin{align}
\widetilde{\bm{E}} &=  \sum _{n=0}^{\infty} \left(\frac{\hbar}{2m}\right)^{2n} \frac{(-1)^{n}}{(2n+1)!}  \left(\partial ^{2n}_{i_{1} ... i_{2n}} \bm{E} \right)  \partial ^{2n}_{v_{i_{1}} \cdots v_{i_{2n}}}  = \bm{E} - \frac{\hbar^{2}}{12 m^{2}} \frac{\partial ^{2} \bm{E}} {x_{i_{1}} ... x_{i_{2}}}  \frac{\partial ^{2}}{\partial_{v_{i_{1}} \cdots v_{i_{2}}}} + \mathcal{O}\left(\hbar^{4} \right), \\
\widetilde{\bm{B}} &=  \sum _{n=0}^{\infty} \left(\frac{\hbar}{2m}\right)^{2n} \frac{(-1)^{n}}{(2n+1)!}  \left(\partial ^{2n}_{i_{1} ... i_{2n}} \bm{B} \right)  \partial ^{2n}_{v_{i_{1}} \cdots v_{i_{2n}}}  = \bm{B} - \frac{\hbar^{2}}{12 m^{2}} \frac{\partial ^{2} \bm{B}} {x_{i_{1}} ... x_{i_{2}}}  \frac{\partial ^{2}}{\partial_{v_{i_{1}} \cdots v_{i_{2}}}} + \mathcal{O}\left(\hbar^{4} \right), \\
\bm{ \Delta \widetilde{\pi}} &=  m\mu_{b} \sum _{n=0}^{\infty} \left(\frac{\hbar}{2m}\right)^{2n+1} \frac{(-1)^{n}(2n+2)}{(2n+3)!}  \left(\partial ^{2n+1}_{i_{1} ... i_{2n+1}} \bm{B} \right)  \partial ^{2n+1}_{v_{i_{1}} \cdots v_{i_{2n+1}}}  =  \frac{\mu_{b} \hbar}{6 } \frac{\partial \bm{B}} {x_{i}}  \frac{\partial }{\partial_{v_{i}}} + \mu_{b}\mathcal{O}\left(\hbar^{3} \right).
\end{align}
From these semiclassical expansions, we notice that the velocity shift $\bm{ \Delta \widetilde{\pi}}$ has a purely quantum origin because the leading term in the expansion is of first order in $\hbar$. Therefore it has no classical counterpart. In the case of $\widetilde{\bm{E}}$ and $\widetilde{\bm{B}}$, the leading term in the expansion simply corresponds to the classical electric or magnetic field.

At zeroth order, the equations for $f_{0}$ and $f_{i}$ decouple, so that one can study the particle motion irrespective of the spin degrees
of freedom. To first order in $\hbar$, Eqs. \eqref{wigner equation f0 condensee avec spin avec B}-\eqref{wigner equation f condensee avec spin avec B} become
\begin{align}
&\frac{\partial f_{0}}{\partial t}
+
\bm{v} \cdot \nabla f_{0}  - \frac{e}{m} \left( \bm{E} + \bm{v} \times \bm{B} \right) \cdot \nabla_{\bm{v}} f_{0} + \frac{\mu_{B} }{2mc^{2}} \left( \bm{E} \times \nabla \right)_{i} f_{i} \nonumber \\
&~
- \frac{\mu_{B}}{m} \nabla \left[ B_{i} - \frac{1}{2c^{2}}\left( \bm{v} \times \bm{E} \right)_{i} \right] \cdot \nabla_{\bm{v}} f_{i}  - \frac{\mu_{B} e}{2m^{2}c^{2}} \left[ \bm{E} \times \left( \bm{B} \times \nabla_{\bm{v}} \right) \right]_{i} f_{i} = 0,
\label{vlasov equation f0 avec spin orbit} \\ \nonumber \\
&
\frac{\partial f_{i}}{\partial t}
+
\bm{v} \cdot \nabla f_{i}  - \frac{e}{m} \left( \bm{E} + \bm{v} \times \bm{B} \right) \cdot \nabla_{\bm{v}} f_{i} + \frac{\mu_{B} }{2mc^{2}} \left( \bm{E} \times \nabla \right)_{i} f_{0} \nonumber \\
&~
- \frac{\mu_{B}}{m} \nabla \left[ B_{i} - \frac{1}{2c^{2}}\left( \bm{v} \times \bm{E} \right)_{i} \right] \cdot \nabla_{\bm{v}} f_{0}  - \frac{\mu_{B} e}{2m^{2}c^{2}} \left[ \bm{E} \times \left( \bm{B} \times \nabla_{\bm{v}} \right) \right]_{i} f_{0}  \nonumber \\
&~
- \frac{2\mu_{B}}{\hbar} \left\{ \left[ \bm{B} - \frac{1}{2 c^{2}} \left(\bm{v} \times \bm{E} \right) \right] \times \bm{f} \right\}_{i} = 0.
\label{vlasov equation f avec spin orbit}
\end{align}
where the factor $\hbar$ is hidden in the definition of the Bohr magneton  $\mu_{B} = e \hbar /(2m)$. The quantum corrections in Eqs. \eqref{vlasov equation f0 avec spin orbit}-\eqref{vlasov equation f avec spin orbit} couple the orbital and the spin dynamics through the Zeeman and spin-orbit interactions. There are no quantum corrections to the orbital electronic dynamics because they would appear only at the second order in $\hbar$. For instance, the Darwin term would not appear in the above equations (even if we had included in the full Wigner equations) because it is a correction of order $\hbar^{2}$ to the orbital motion of the electron. In summary,  Eqs. \eqref{vlasov equation f0 avec spin orbit}-\eqref{vlasov equation f avec spin orbit} represent a semiclassical model where the orbital dynamics is classical (hence the familiar Lorentz force terms), while the spin is treated as a fully quantum variable (two-dimensional Hilbert space).

In Eqs \eqref{vlasov equation f0 avec spin orbit} and \eqref{vlasov equation f avec spin orbit}, the Zeeman interaction gives two contributions: (i)  The term $\mu_{b} \nabla B_{i} \cdot \nabla_{\bm{v}}$, which represents the force exerted on a magnetic dipole by an inhomogeneous magnetic field and is at the basis of Stern-Gerlach-type experiments; and (ii) the term $\bm{f} \times \bm{B}$  which describes the precession of the spin around the magnetic field lines.

The spin-orbit interaction yields a correction to the magnetic field $\bm{B} \rightarrow \bm{B} - \left(\bm{v} \times \bm{E} \right)/2c^{2}$, which is the first-order correction in the nonrelativistic limit of the Thomas precession \cite{thomas original,thomas new}. The other terms are related to the spin-orbit  correction of the velocity operator. Indeed, in the Heisenberg picture, the velocity operators $ \bm{\widehat{\mathit{V}}}$ is determined by the evolution equation of the position operator
\begin{align}
\widehat{\bm{\mathit{V}}} &= \frac{1}{i\hbar} \left[ \bm{\widehat{R}} , \widehat{\mathcal{H}} \right] = \frac{\bm{\widehat{\pi}}}{m} -\frac{\mu_{b}}{ 2mc^{2}} \bm{\widehat{E}} \times \bm{\sigma},
\label{velocity operators}
\end{align}
where we used the Hamiltonian defined in  Eq. \eqref{Pauli equation}. The associated phase space function is determined by the Weyl correspondence rule and reads as
\begin{align}
\bm{\mathit{V}} = \bm{v} - \frac{\mu_{b}}{2mc^{2}} \bm{E} \times \bm{\sigma}.
\label{velocity phase space function}
\end{align}
This is the phase space function that can be used to calculate the average velocity or the charge current. Therefore, the particles are transported with a modified velocity. The term $\left( \bm{E} \times \nabla \right)_{i}$ in Eq. \eqref{vlasov equation f0 avec spin orbit}-\eqref{vlasov equation f avec spin orbit} is a direct consequence of this effect, whereas the term $\left[ \bm{E} \times \left( \bm{B} \times \nabla_{\bm{v}} \right) \right]_{i}$ corresponds to the same velocity correction in the Lorentz force $ \bm{v} \times \bm{B}$.

The spin Vlasov equations \eqref{vlasov equation f0 avec spin orbit}-\eqref{vlasov equation f avec spin orbit} are correct to second order in $1/c$ and can be viewed as are a generalisation of those obtained in Ref. \cite{hurst}, where only the Zeeman interaction was included. An alternative form of these equations was obtained by Asenjo et al. \cite{asenjo_kinetic_so} in the extended phase space formalism.

The Maxwell equations \eqref{maxwell equations}, combined with the spin Vlasov equations \eqref{vlasov equation f0 avec spin orbit}-\eqref{vlasov equation f avec spin orbit}, form a self-consistent model for the charge and the spin dynamics of a system of interacting particles. One can show that the following quantities are conserved during the time evolution:
\begin{align}
M_{tot}   &= m \int f_{0} d\bm{v}d\bm{r}, \\
\bm{P}_{tot} &= m  \int  v f_{0}  d\bm{v}d\bm{r} +  \int \bm{D} \times \bm{B} d\bm{r} , \\
E_{tot} &=\frac{m}{2}  \int  \bm{v}^{2} f_{0} d\bm{v}d\bm{r}  + \mu_{B} \int \bm{f} \cdot \bm{B}d\bm{v}d\bm{r} + \frac{\epsilon_{0}}{2} \int \bm{E}^{2} d\bm{r} + \frac{1}{2 \mu_{0}} \int \bm{B}^{2} d\bm{r}, \\
\bm{J}_{tot} &= m \int \left( \bm{r} \times \bm{v} \right) f_{0} d\bm{r} d\bm{v} +  \frac{\hbar}{2} \int \bm{f} d\bm{r} d\bm{v} + \int \bm{r} \times \left( \bm{D} \times \bm{B} \right) d\bm{r},
\end{align}
where $\bm{D} = \epsilon_{0}\bm{E} + \bm{P}$ and $\bm{H} = \bm{B} - \mu_{0} \bm{M}$. The conserved quantities are the total mass $M_{tot}$, the total linear momentum $\bm{P}_{tot}$ (sum of the particles and fields momenta), the total energy $E_{tot}$ (sum of the kinetic, Zeeman, and the electromagnetic field energies), and the total angular momentum $\bm{J}_{tot}$ (sum of the orbital, spin, and electromagnetic field angular momenta).

For simulation purposes, the spin Vlasov equations \eqref{vlasov equation f0 avec spin orbit}-\eqref{vlasov equation f avec spin orbit} are much easier to solve numerically than the corresponding Wigner equations \eqref{wigner equation f0 condensee avec spin avec B}-\eqref{wigner equation f condensee avec spin avec B}, mainly because the former are local in space while the latter are not. The Vlasov approximation is valid when quantum effects in the orbital dynamics are small. From Eq. \eqref{Moyal bracket}, it appears that the semiclassical expansion is valid when $\hbar/(mL_{0}v_{0}) \ll 1$, where $L_{0}$ and $v_{0}$ are typical length and velocity scales. For a degenerate electron gas with density $n$, the typical velocity is the Fermi speed $v_F = \hbar (3\pi^2n)^{1/3}/m$. Inserting into the previous inequality, we obtain  the validity condition $L_0 n^{1/3} \gg 1$, which means that the typical length scale must be larger than the interparticle distance $d= n^{-1/3}$. For this reason, the semiclassical limit is also referred to as the long wavelength approximation.
All in all, the above spin Vlasov equations constitute a valuable tool to simulate the charge and spin dynamics in condensed matter systems, particularly semiconductor and metallic nano-objects.

\section{Hydrodynamic model with spin-orbit coupling}

In this Section, starting from Eqs. \eqref{wigner equation f0 condensee avec spin avec B}-\eqref{wigner equation f condensee avec spin avec B}, we derive the hydrodynamic evolution equations by taking velocity moments of the phase-space distribution functions.
In addition to the particle density and spin polarization [Eqs.
\eqref{def n} and \eqref{def S}], we define the following macroscopic quantities
\begin{align}
\bm{u} &= \frac{1}{n}\int \bm{v} f_{0}d\bm{v},\label{def u} \\
J^{S}_{i\alpha}&= \int v_{i} f_{\alpha}d\bm{v},\label{def J}\\
P_{ij}&= m\int w_{i} w_{j} f_{0}d\bm{v},\label{def P}\\
\Pi_{ij\alpha} &= m  \int v_{i} v_{j}  f_{\alpha}  d\bm{v},\label{def Pi}\\
Q_{ijk} &= m \int w_{i} w_{j} w_{k} f_{0} d\bm{v},\label{def Q}
\end{align}
where we separated the mean fluid velocity $\bm u$ from the velocity fluctuations $\bm{w} \equiv \bm{v} - \bm{u}$.
Here, $P_{ij}$ and $Q_{ijk} $ are respectively the pressure and the generalized energy flux tensors. They coincide with the analogous definitions for spinless fluids with probability distribution function $f_0$.
The spin-velocity tensor $J_{i \alpha}^S$ represents  the mean fluid velocity along the $i$-th direction of the $\alpha$-th spin polarization vector, while $\Pi_{ij\alpha}$ represents the corresponding spin-pressure tensor
\footnote{Strictly speaking a pressure tensor should be defined in terms of the velocity fluctuations $w_i w_j$, but this would unduly complicate the notation. Thus, we stick to the above definition of $\Pi_{ij\alpha}$ while still using the term ``pressure" for this quantity.}.
The evolution equations for the above fluid quantities are obtained by the straightforward integration of Eqs. \eqref{vlasov equation f0 avec spin orbit}-\eqref{vlasov equation f avec spin orbit} with respect to the velocity variable. One obtains :
\begin{align}
&\frac{\partial n}{\partial t} + \bm{\nabla}_{\bm{r}} \cdot \left(n\bm{\overline{u}}\right)   = 0,  \label{f0_continuity} \\
&\frac{\partial S_{\alpha}}{\partial t} + \partial_{i} \overline{J}^{S}_{i \alpha} + \frac{e}{ m } \left( \bm{S} \times \bm{B} \right) _{\alpha} +  \frac{e}{2mc^{2}} \epsilon_{jk\alpha} \epsilon_{rlj} E_{l} J^{S}_{rk}=  0, \label{falpha_continuity} \\
&\frac{\partial u_{i} }{\partial t} + u_{j} (\nabla_{j} u_{i}) + \frac{1}{nm} \nabla_{j} P_{ij} + \frac{e}{m} \left[E_{i} + \left(\bm{\widetilde{u}} \times \bm{B} \right)_{i}  \right]+\frac{e}{nm^{2}}  S_{\alpha} \left(\partial_{i} B_{\alpha} \right) \nonumber \\
&~
+\frac{\mu_{b}}{2mc^{2} n } \epsilon_{jkl} \left[ u_{i} \partial_{j} \left( E_{k} S_{l} \right) +  E_{j} \left(\partial_{k} J^{S}_{il}\right) -   \left(\partial_{i} E_{k} \right) J^{S}_{jl}   -   \left(\partial_{j} E_{k} \right) J^{S}_{il} \right]
= 0,  \label{euler eq}\\
&\frac{\partial J^{S}_{i \alpha}}{\partial t} + \partial_{j} \Pi_{ij\alpha} + \frac{e E_{i}}{ m }S_{\alpha} + \frac{e}{ m } \epsilon_{jki} B_{k} \widetilde{J}^{S}_{j\alpha} + \frac{e}{ m } \epsilon_{jk\alpha} B_{k} J^{S}_{ij} + \frac{\mu_{B} \hbar}{2m}  \left(\partial_{i} B_{\alpha} \right) n \nonumber \\
&~
+ \frac{\mu_{b}}{2mc^{2}} \epsilon_{kl\alpha}\partial_{l} \left( E_{k} n u_{i} \right)- \frac{\mu_{b}}{2mc^{2} } \epsilon_{kl\alpha}  \left(\partial_{i} E_{l} \right) n u_{k} + \frac{\mu_{b}}{\hbar c^{2}} \epsilon_{kl\alpha} \epsilon_{rsk} E_{s} \Pi^{S}_{irl} = 0,\label{evol Js}\\
&
\frac{\partial P_{ij}}{\partial t}+ u_{k} \partial_{k} P_{ij} + P_{jk}  \partial_{k} u_{i} + P_{ik} \partial_{k} u_{j}  + P_{ij} \partial_{k} u_{k} +  \partial_{k} Q_{ijk}
+\frac{e}{m} \left[ \epsilon_{kli}  P_{jk}  +  \epsilon_{klj}   P_{ik} \right] B_{l} \nonumber \\
&~
+ \frac{\mu_{b}}{m}\left[ \partial_{i} B_{k}  \left( J^{S}_{jk} - u_{j} S_{k} \right) + \partial_{j} B_{k}  \left( J^{S}_{ik} - U_{i}S_{k} \right) \right]
+ \frac{\mu_{b}}{2mc^{2}} \epsilon_{rsl} \partial_{s} \left[ E_{r}\left( \Pi^{S}_{ijl} - u_{i}u_{j}S_{l} \right)\right] \nonumber \\
&~
+  \frac{\mu_{b}}{2mc^{2}}  \epsilon_{rkp} E_{r} \left[ \epsilon_{kli}   \left(J^{S}_{jp}-u_{j}S_{p} \right) +  \epsilon_{klj}   \left( J^{S}_{ip} - u_{i}S_{p} \right) \right] B_{l}\nonumber \\
&~
- \frac{\mu_{b}}{2mc^{2}} \epsilon_{rsl} \left[  \partial_{i}  E_{s}  \left(\Pi^{S}_{jrl} - u_{j} J^{S}_{rl} \right)  + \partial_{j}  E_{s}  \left(\Pi^{S}_{irl} - u_{i} J^{S}_{rl} \right)  \right]\nonumber \\
&~
-
\frac{\mu_{b}}{2mc^{2} }u_{i} \epsilon_{rsl}  \partial_{s} \left[ E_{r} \left( J^{S}_{jl} -  u_{j} S_{l} \right) \right]
-
\frac{\mu_{b}}{2mc^{2} }u_{j} \epsilon_{rsl}  \partial_{s} \left[ E_{r} \left( J^{S}_{il} -  u_{i} S_{l} \right) \right] =0,
\label{eq_presure}
\end{align}
where we introduced a new average velocity and a new spin current
\begin{align}
 \overline{\bm{u}}  = \bm{u}- \frac{\mu_{b}}{2mc^{2}n} \bm{E} \times \bm{S},~~~~
\overline{J}^{S}_{ij} = J^{S}_{ij}+ \frac{\mu_{b}}{2mc^{2}} \epsilon_{ijk} E_{k}n.
\end{align}

The above corrections reflect the modification of the velocity due to the spin-orbit coupling. Indeed the average velocity can be immediately obtained from the velocity phase space function, Eq. \eqref{velocity phase space function}, yielding
\begin{align}
 \overline{\bm{u}}  &= \frac{1}{nm} \textrm{tr} \left[ \int \bm{\mathit{V}}(\bm{r},\bm{\pi}) F d\bm{\pi}  \right]  = \bm{u}- \frac{\mu_{b}}{2mc^{2}n} \bm{E} \times \bm{S},
\label{spin orbit correction velocity}
\end{align}
where $F$ is the $2 \times 2$ distribution function defined in Eq. \eqref{change basis wigner function}.
The same holds for the spin current operator, which is defined as follows:
\begin{align}
\widehat{J}^{S}_{ij} = \widehat{v}_{i}\sigma_{j} = \frac{\widehat{\pi}_{i}}{m} \sigma_{j} - \frac{\mu_{b}}{4mc^{2}} \left[ \left( \widehat{\bm{E}}\times \bm{\sigma} \right)_{i} \sigma_{j} + \sigma_{j} \left( \widehat{\bm{E}} \times \bm{\sigma} \right)_{i} \right],
\end{align}
where we symmetrized the operator so that it is Hermitian. Then the associated phase-space function
\begin{align}
J^{S}_{ij} \left( \bm{r} , \bm{\pi}\right)  = \frac{\pi_{i}}{m} \sigma_{j} - \frac{\mu_{b}}{4mc^{2}} \left[ \left( \bm{E}\times \bm{\sigma} \right)_{i} \sigma_{j} + \sigma_{j} \left( \bm{E} \times \bm{\sigma} \right)_{i} \right]
\end{align}
can be used to determine the spin current
\begin{align}
 \overline{J}^{S}_{ij} = \frac{1}{m} \textrm{tr} \left[ \int J^{s}_{ij}(\bm{r},\bm{\pi}) F d\bm{\pi}  \right]  = J^{S}_{ij}+ \frac{\mu_{b}}{2mc^{2}} \epsilon_{ijk} E_{k}n.
\end{align}

As is always the case for hydrodynamic models, some further hypotheses are needed to close the above set of equations \eqref{f0_continuity}-\eqref{eq_presure}. A particularly interesting strategy, based on the maximum entropy method (MEP), was used in a previous work \cite{hurst} to close the set of hydrodynamic equations in the presence of the sole Zeeman interaction. Unfortunately, when one adds the spin orbit interaction, the MEP  does not provide any conclusive analytical results (the difficulty arises from the fact that the spin-orbit interaction couples all the components of the velocity). However an intuitive closure can be found by inspecting the evolution equation \eqref{eq_presure} for the pressure tensor. There, most spin-dependent terms cancel if we set
\begin{align}
J^{S}_{i\alpha} = u_{i} S_{\alpha} ~~~~\textrm{and}~~~~\Pi^{S}_{ij\alpha} = u_{i} J^{S}_{j\alpha}.  \label{int clos}
\end{align}
The physical interpretation of the above equations is that the spin of a particle is simply transported along the mean fluid velocity.
This is of course an approximation that amounts to neglecting some spin-velocity correlations \cite{zamanian_POP10}.
With this assumption,  Eq. \eqref{evol Js} and the definition of the spin-pressure $\Pi_{ij\alpha}$ are no longer necessary. The system of fluid equations simplifies to
\begin{align}
&\frac{\partial n}{\partial t} + \bm{\nabla}_{\bm{r}} \cdot \left(n\bm{\overline{u}}\right)   = 0,  \label{f0_continuity closed} \\
&\frac{\partial S_{\alpha}}{\partial t} + \partial_{i}\left(u_{i} S_{\alpha}\right) - \frac{\mu_{b}}{2mc^{2}} \left( \bm{\nabla} \times n\bm{E} \right)_{\alpha}  + \frac{e}{ m } \left[ \bm{S} \times \left(\bm{B} - \frac{1}{2c^{2}} \bm{u} \times \bm{E} \right) \right] _{\alpha} =  0, \label{falpha_continuity closed} \\
&\frac{\partial u_{i} }{\partial t} + u_{j} (\nabla_{j} u_{i}) + \frac{1}{nm} \nabla_{j} P_{ij} + \frac{e}{m} \left[E_{i} + \left(\bm{\overline{u}} \times \bm{B} \right)_{i}  \right]+\frac{e}{nm^{2}}  S_{\alpha} \left(\partial_{i} B_{\alpha} \right) \nonumber \\
&~
+\frac{\mu_{b}}{2mc^{2} n } \epsilon_{jkl} \left[ E_{j} \left(\partial_{k}u_{i}\right) - u_{k} \left( \partial_{i} E_{j} \right)  \right]S_{l}
= 0,  \label{euler eq closed}\\
&
\frac{\partial P_{ij}}{\partial t}+ u_{k} \partial_{k} P_{ij} + P_{jk}  \partial_{k} u_{i} + P_{ik} \partial_{k} u_{j}  + P_{ij} \partial_{k} u_{k} +  \partial_{k} Q_{ijk}
+\frac{e}{m} \left[ \epsilon_{kli}  P_{jk}  +  \epsilon_{klj}   P_{ik} \right] B_{l}=0.
\label{eq_presure closed}
\end{align}
In order to complete the closure procedure, one can proceed in the same way as is usually done for spinless fluids, for instance by supposing that the system is isotropic and adiabatic.
The isotropy condition imposes that $P_{ij} = (P/3) \delta_{ij}$ where $\delta_{ij}$ is the Kronecker delta, while the adiabaticity condition requires that the heat flux $Q_{ijk}$ vanish. In that case, the pressure takes the usual form of the equation of state of an adiabatic system, i.e.,
$P ={\rm const.} \times n^{\frac{D+2}{D}}$
(where $D$ is the dimensionality of the system), which replaces Eq. \eqref{eq_presure closed}.
In summary, Eqs. \eqref{f0_continuity closed}-\eqref{euler eq closed}, together with the preceding expression for the pressure, constitute a closed system of hydrodynamic equations with spin-orbit effects.

\section{Conclusions}

Phase space methods can be applied to condensed-matter and nanophysics to model the electron dynamics either in the quantum or the semiclassical regime. Several studies were performed in the past but neglected spin effects \cite{Calvayrac,metal_films_2,metal_films_3}.
In this paper, we show that phase-space methods can be conveniently generalized to include the spin dynamics at different orders. In an earlier  work, we had developed a phases-space model that includes the lowest order spin term (the Zeeman effect), but neglects all relativistic corrections (spin-orbit coupling, Darwin term, mass correction, \dots).
Here, we considered the case where both the Zeeman and the spin-orbit interaction are present (other relativistic corrections could be added with relative ease). The spin-orbit interaction plays an important role, for instance, in ultrafast spectroscopy experiments on magnetic nano-objects, where the electron spin is known to interact with the incident laser field and with the self-consistent field generated by the electron gas.

We first derived a four-component Wigner equation to describe the quantum dynamics of a system of spin-1/2 particles.
These equations, together with the appropriate Maxwell equations, form a fully quantum self-consistent model to study the spin and charge dynamics in the mean field approximation. This model is not limited to the linear response, but can deal with nonlinear effects, which are often important, particularly for large incident laser powers.

Next, using a semiclassical expansion to first order in $\hbar$, we obtained a four-component Vlasov equation. The orbital part of the motion is classical, i.e., the particles follow the classical phase-space trajectories, while the spin degrees of freedom are treated in a fully quantum fashion (two-dimensional Hilbert space). These spin Vlasov equations constitute a good approximation
of the quantum electron dynamics for wavelengths larger than the typical inter-electron distance.

The corresponding hydrodynamic equations were derived by taking velocity moments of the phase-space distribution functions. The spin-orbit interaction modifies considerably our earlier hydrodynamic equations \cite{hurst}, where the only spin effect was the Zeeman interaction. We proposed a simple, intuitive closure for the hydrodynamic equations whereby the spin is simply transported along by the fluid velocity of the electrons.

The present models (Vlasov and hydrodynamic) constitute two valuable tools to study the intertwined spin and charge dynamics in condense-matter systems and nano-objects. The challenge now is to implement these models into performing numerical codes and to study the electron dynamics in realistic nanoscale systems that are relevant to current experiments.

\vskip 0.5cm
{\it \noindent Acknowledgments}\\
We thank the Agence Nationale de la Recherche, project Labex ``Nanostructures in Interaction with their Environment", for financial support.

\newpage

\input{supplementary2.tex}

\end{document}

%% file: supplementary2.tex
\appendix
\section{Supplementary Material}
\begin{center}
{\bf \large DERIVATION OF THE SPIN WIGNER EQUATIONS}
\end{center}
\vskip 5mm

\noindent
The evolution equation of the density matrix in the case of an electron interacting with an electromagnetic field reads as
\begin{equation}
i \hbar \frac{\partial \widehat{\rho}}{\partial t} = \left[ \widehat{H} , \widehat{\rho}\right], ~~~~\textrm{with}~~~~
\widehat{\rho} =
\begin{pmatrix}
\widehat{\rho}^{\uparrow \uparrow} & \widehat{\rho}^{\uparrow \downarrow}\\
\widehat{\rho}^{\downarrow \uparrow } & \widehat{\rho}^{\downarrow \downarrow}
\end{pmatrix}
~~~~\textrm{and}~~~~
\widehat{H} =
\begin{pmatrix}
\widehat{H} ^{\uparrow \uparrow} & \widehat{H} ^{\uparrow \downarrow}\\
\widehat{H} ^{\downarrow \uparrow } & \widehat{H} ^{\downarrow \downarrow}
\end{pmatrix}.
\end{equation}
In the forthcoming derivation, we shall include the Zeeman and the spin-orbit interaction. The four components of the Hamiltonian are as follows
\begin{align}
\widehat{H} ^{\uparrow \uparrow}
&=
\frac{\widehat{\bm{\pi}}^{2}} {2m}  + V(\widehat{\bm{R}}) + \mu_{B} B_{z} (\widehat{\bm{R}} ) + \frac{\mu_{b}}{4mc^{2}} \left[ E(\widehat{\bm{R}})  \times \widehat{\bm{\pi}}  - \widehat{\bm{\pi}} \times E(\widehat{\bm{R}})\right]_{z}
\label{operateur h up up}, \\ \nonumber \\
\widehat{H} ^{\uparrow \downarrow}
&=
\frac{\mu_{b}}{4mc^{2}} \left\{ \left[  E(\widehat{\bm{R}})  \times \widehat{\bm{\pi}}  - \widehat{\bm{\pi}} \times E(\widehat{\bm{R}}) \right]_{x} - i \left[  E(\widehat{\bm{R}})  \times \widehat{\bm{\pi}}  - \widehat{\bm{\pi}} \times E(\widehat{\bm{R}}) \right]_{y} \right\}
\label{operateur h up down}, \\ \nonumber \\
\widehat{H} ^{\downarrow \uparrow}
&=
\frac{\mu_{b}}{4mc^{2}} \left\{ \left[  E(\widehat{\bm{R}})  \times \widehat{\bm{\pi}}  - \widehat{\bm{\pi}} \times E(\widehat{\bm{R}}) \right]_{x} + i \left[  E(\widehat{\bm{R}})  \times \widehat{\bm{\pi}}  - \widehat{\bm{\pi}} \times E(\widehat{\bm{R}}) \right]_{y} \right\}
\label{operateur h down up}, \\ \nonumber \\
\widehat{H} ^{\downarrow \downarrow}
&=
\frac{\widehat{\bm{\pi}}^{2}} {2m}  + V(\widehat{\bm{R}}) - \mu_{B} B_{z} (\widehat{\bm{R}} ) - \frac{\mu_{b}}{4mc^{2}} \left[ E(\widehat{\bm{R}})  \times \widehat{\bm{\pi}}  - \widehat{\bm{\pi}} \times E(\widehat{\bm{R}})\right]_{z}
\label{operateur h down down}
\end{align}
In order to compute the phase space function corresponding to the above Hamiltonian, we apply the Weyl transformation, described in Sec. II of the main paper. Here, we shall give the details of the Weyl transformation for the first term $\widehat{H} ^{\uparrow \uparrow}$.
First, we should write the Hamiltonian \eqref{operateur h up up} in a symmetric form
\begin{align}
\widehat{H} ^{\uparrow \uparrow}
&=
\frac{\widehat{\bm{\pi}}^{2}} {2m}  + V(\widehat{\bm{R}}) + \mu_{B} B_{z} (\widehat{\bm{R}} ) + \frac{\mu_{b}}{4mc^{2}} \epsilon_{ijz} \left(\frac{ E_{i}(\widehat{\bm{R}})  \widehat{\pi_{j}}}{2} + \frac{  \widehat{\pi_{j}} E_{i}(\widehat{\bm{R}})}{2} + \frac{i\hbar}{2} \partial_{j}E_{i} \right) \nonumber \\
&~~~
-\frac{\mu_{b}}{4mc^{2}} \epsilon_{ijz} \left(\frac{ \widehat{\pi_{i}} E_{j}(\widehat{\bm{R}})}{2} + \frac{   E_{j}(\widehat{\bm{R}})\widehat{\pi_{i}}}{2}  - \frac{i\hbar}{2} \partial_{i}E_{j} \right) \nonumber \\
&=
\frac{\widehat{\bm{\pi}}^{2}} {2m}  + V(\widehat{\bm{R}}) + \mu_{B} B_{z} (\widehat{\bm{R}} ) + \frac{\mu_{b}}{4mc^{2}} \epsilon_{ijz} \left(\frac{ E_{i}(\widehat{\bm{R}})  \widehat{\pi_{j}}}{2} + \frac{  \widehat{\pi_{j}} E_{i}(\widehat{\bm{R}})}{2} - \frac{ \widehat{\pi_{i}} E_{j}(\widehat{\bm{R}})}{2} - \frac{   E_{j}(\widehat{\bm{R}})\widehat{\pi_{i}}}{2}  \right), \nonumber \\
\end{align}
where we used the following commutation relation: $\left[ \widehat{\pi_{i}} , F  (\widehat{\bm{R}} ) \right] = -i \hbar \partial_{i} F$. Then we replace all operators with their associated phase space variables. By doing so, we obtain the phase space function associated to $\widehat{H} ^{\uparrow \uparrow}$:
\begin{align}
H ^{\uparrow \uparrow}
&=
 \frac{\bm{\pi}^{2}} {2m}  + V + \mu_{B} B_{z}  + \frac{\mu_{b}}{2mc^{2}} \left[ \bm{E} \times \bm{\pi} \right]_{z}.
 \label{weyl h up up}
\end{align}
The phase space function for the other components of the Hamiltonian can be obtained through similar calculations:
\begin{align}
H ^{\uparrow \downarrow}
&=
\mu_{b} \left(B_{x} - i B_{y} \right)
+
\frac{\mu_{b}}{2mc^{2}} \left( \left[ \bm{E} \times \bm{\pi}  \right]_{x} - i \left[\bm{E} \times \bm{\pi} \right]_{y} \right)
\label{weyl h up down}, \\ \nonumber \\
H^{\downarrow \uparrow}
&=
\mu_{b} \left(B_{x} + i B_{y} \right)
+
\frac{\mu_{b}}{2mc^{2}} \left( \left[ \bm{E} \times \bm{\pi} \right]_{x} + i \left[  \bm{E} \times \bm{\pi}  \right]_{y} \right)
\label{weyl h down up}, \\ \nonumber \\
H ^{\downarrow \downarrow}
&=
\frac{\bm{\pi}^{2}} {2m}  + V - \mu_{B} B_{z}  - \frac{\mu_{b}}{2mc^{2}} \left[ \bm{E} \times \bm{\pi} \right]_{z}.
\label{weyl h down down}
\end{align}
The equations of motion for the four Wigner functions are determined by Eqs. (5)-(6) in the main text. Using these equations, one obtains:
\begin{align}
i \hbar \partial_{t} f_{0}
&=
\left[\frac{\bm{\pi}^{2}} {2m}  + V , f_{0} \right]_{\star} + \mu_{B} \left[ B_{i},f_{i} \right]_{\star} +  \frac{\mu_{b}}{2mc^{2}} \epsilon_{ijk} \left[  E_{i} \pi_{j} , f_{k} \right]_{\star}, \label{eq moyal f0} \\
i \hbar \partial_{t} f_{k}
&=
\left[\frac{\bm{\pi}^{2}} {2m}  + V , f_{k} \right]_{\star} + \mu_{B} \left[ B_{k},f_{0} \right]_{\star} +  \frac{\mu_{b}}{2mc^{2}} \epsilon_{ijk} \left[  E_{i} \pi_{j} , f_{0} \right]_{\star} + i \mu_{B} \epsilon_{ijk} \left\{ B_{i},f_{j} \right\}_{\star}\nonumber \\
&~~~
 + i  \frac{\mu_{b}}{2mc^{2}}  \epsilon_{lri} \epsilon_{ijk} \left\{  E_{l} \pi_{r} , f_{j} \right\}_{\star},
\label{eq moyal f}
\end{align}
where $\left\{,\right\}$ denotes the anti-commutator. From the Eqs. (11)-(12) in the main text, we notice that the operators $\mathcal{L}$ and $\mathcal{L}_{n}$ commute with each other, so we can rewrite the Moyal product [Eq. (10) in the main text] in a more convenient way for the calculations:
\begin{align}
A(\bm{r},\bm{\pi}) \star C(\bm{r},\bm{\pi}) &= \prod_{n=1}^{\infty} \exp \left(  ie \sum_{n=1}^{\infty} \hbar^{n} \mathcal{L}_{n} \right) \exp \left( i \hbar \mathcal{L} \right)\left( A(\bm{r},\bm{\pi}), C(\bm{r},\bm{\pi})\right).
\label{moyal product l ln factorized}
\end{align}
We will also use the following symmetry properties for the operators  $\mathcal{L}$ and $\mathcal{L}_{n}$
\begin{align}
\mathcal{L}^{n} \left(A,B\right) &= \left(-1\right)^{n}\mathcal{L}^{n} \left(B,A\right),~~~~\mathcal{L}_{n}^{m} \left(A,B\right) = \left(-1\right)^{nm}\mathcal{L}_{n}^{m} \left(B,A\right).
\label{symmetric properties of l ln}
\end{align}
In order to develop Eqs. \eqref{eq moyal f0} and \eqref{eq moyal f}, we calculate separately the following five terms $\left[\frac{\bm{\pi}^{2}} {2m} , f_{0} \right]_{\star}$, $\left[V  , f_{0} \right]_{\star}$, $\left[  E_{i} \pi_{j} , f_{k} \right]_{\star}$, $\left\{ B_{i},f_{j} \right\}_{\star}$ and $\left\{  E_{i} \pi_{j} , f_{k} \right\}_{\star}$.

\begin{enumerate}

\item Term: $\left[\frac{\bm{\pi}^{2}} {2m}  , f_{0} \right]_{\star}$.

From Eq. (10) in the main text, the above commutator reads as
\begin{align}
\left[\frac{\bm{\pi}^{2}} {2m}  , f_{0} \right]_{\star}  &=
\exp \left[ i \hbar \mathcal{L} + ie \sum_{n=1}^{\infty} \hbar^{n} \mathcal{L}_{n} \right] \left[ \left(\frac{\bm{\pi}^{2}} {2m}  , f_{0}\right)- \left(f_{0}, \frac{\bm{\pi}^{2}} {2m} \right) \right]
\label{pi2f0}
\end{align}
Let us develop the following quantity:
\begin{align}
&
\exp \left( i \hbar \mathcal{L} \right) \left( f(\bm{r},\bm{\pi}) ,\frac{\bm{\pi}^{2}} {2m}\right)
=
\sum_{n=0}^{\infty} \left( \frac{i \hbar}{2}\right)^{n} \frac{1}{n!}  \left( {}^L \! \partial_{i} {}^R \! \partial_{\pi_{i}} - {}^R \! \partial_{j} {}^L \! \partial_{\pi_{j}}  \right)^{n} \left( f ,\frac{\bm{\pi}^{2}} {2m}\right)\nonumber \\
&=  \sum_{n=0}^{\infty}  \sum_{p=0}^{n}  \left(\frac{i\hbar}{2}\right)^{p} \frac{(-1)^{p}}{n!}
\begin{pmatrix}
n\\p
\end{pmatrix}
 \left(  \partial_{i_{1} \cdots i_{n-p}}^{n-p} \partial_{\pi_{j_{1}} \cdots \pi_{j_{p}}}^{p} f \right)  \left( \partial_{\pi_{i_{1}} \cdots \pi_{i_{n-p}}}^{n-p} \partial_{j_{1} \cdots j_{p}}^{p} \frac{\bm{\pi}^{2}} {2m}\right) \nonumber .
\end{align}
This expression differs from zero only if $p=0$, so one obtains:
\begin{align}
\exp \left(i \hbar \mathcal{L} \right) f(\bm{r},\bm{\pi}) H_{\mathcal{W}} (\bm{\pi})
&=
\sum_{n=0}^{\infty}  \left(\frac{i\hbar}{2}\right)^{n} \frac{1}{n!}
\left( \partial_{\pi_{i_{1}} \cdots \pi_{i_{n}}}^{n} \frac{\bm{\pi}^{2}}{2m} \right) \partial_{i_{1} \cdots i_{n}}^{n} f, \nonumber \\
&=
 \frac{\bm{\pi}^{2}}{2m} f + \frac{i \hbar}{2m} \pi_{i} \left(\partial_{i} f\right) - \frac{\hbar^{2}}{8m} \sum_{i=1}^{3}\left( \partial_{i}^{2} f\right).
 \label{exp L sur H f}
\end{align}
Using the symmetry properties \eqref{symmetric properties of l ln} for the operator $\mathcal{L}$, one obtains:
\begin{align}
\exp \left( i \hbar \mathcal{L} \right)   \left(\frac{\bm{\pi}^{2}} {2m}, f(\bm{r},\bm{\pi})\right)
&=
 \frac{\bm{\pi}^{2}}{2m} f - \frac{i \hbar}{2m} \pi_{i} \left(\partial_{x_{i}} f\right) - \frac{\hbar^{2}}{8m} \sum_{i=1}^{3}\left( \partial_{x_{i}}^{2} f\right).
\label{exp L sur f H}
\end{align}
Injecting Eqs. \eqref{exp L sur H f} and  \eqref{exp L sur f H} in Eq. \eqref{pi2f0} gives
\begin{align}
\left[\frac{\bm{\pi}^{2}} {2m}  , f_{0} \right]_{\star} &= \frac{1}{2m}\prod_{n=1}^{\infty} \exp \left(  ie \sum_{n=1}^{\infty} \hbar^{n} \mathcal{L}_{n} \right) \left(   \pi_{k}^{2},f\right)  - \frac{1}{2m}\prod_{n=1}^{\infty} \exp \left(  ie \sum_{n=1}^{\infty} \hbar^{n} \mathcal{L}_{n} \right) \left(  f,\pi_{k}^{2}\right)  \nonumber \\
&~  - \frac{i \hbar}{2m}\prod_{n=1}^{\infty} \exp \left(  ie \sum_{n=1}^{\infty} \hbar^{n} \mathcal{L}_{n} \right) \left(   \pi_{i},\partial_{i} f\right) - \frac{i \hbar}{2m}\prod_{n=1}^{\infty} \exp \left(  ie \sum_{n=1}^{\infty} \hbar^{n} \mathcal{L}_{n} \right) \left( \partial_{i} f,\pi_{i}\right)
\label{developpement commutateur moyal factorisation Ln}
\end{align}
From Eq. (12) in the main text, the $\mathcal{L}_{n}$ operator contains at least one derivative in $\bm{\pi}$. Then, by developing the exponentials in power series in Eq. \eqref{developpement commutateur moyal factorisation Ln},  all the operators of order higher than $n=2$ will give no contributions. Then Eq. \eqref{developpement commutateur moyal factorisation Ln} reduces to
\begin{align}
\left[\frac{\bm{\pi}^{2}} {2m}  , f_{0} \right]_{\star}
&= -\frac{i\hbar}{m} \left(\pi_{i} \partial_{x_{i}} f \right) + \frac{1}{2m}\left[ie \sum_{n=1}^{\infty} \hbar^{n} \mathcal{L}_{n} - \frac{e^{2}}{2} \left(\sum_{n=1}^{\infty} \hbar^{n} \mathcal{L}_{n} \right)^{2} \right] \left[\left(   \pi_{k}^{2},f\right) - \left(  f,\pi_{k}^{2}\right) \right]\nonumber \\
&~~~
+ \frac{ e\hbar}{2m}\left[  \sum_{n=1}^{\infty} \hbar^{n} \mathcal{L}_{n} \right]  \left[\left(   \pi_{i},\partial_{i} f\right) + \left( \partial_{i} f,\pi_{i}\right) \right]
\label{developpement commutateur moyal factorisation Ln ordre 1_2}
\end{align}
Using the symmetry properties \eqref{symmetric properties of l ln} of the operator $\mathcal{L}_{n}$, one obtains:
\begin{align}
\left[\frac{\bm{\pi}^{2}} {2m}  , f_{0} \right]_{\star}
&= -\frac{i\hbar}{m} \left(\pi_{i} \partial_{x_{i}} f \right)   + \frac{ e\hbar}{m}  \sum_{n=0}^{\infty} \hbar^{2n+2} \mathcal{L}_{2n+2}   \left(   \pi_{i},\partial_{x_{i}} f\right) + \frac{ie}{m} \sum_{n=0}^{\infty} \hbar^{2n+1} \mathcal{L}_{2n+1} \left(   \pi_{k}^{2},f\right)   \nonumber \\
&~
- \frac{e^{2}}{m} \sum_{n=0}^{\infty} \sum_{p=0}^{\infty} \hbar^{2n+1} \hbar^{2p+2} \mathcal{L}_{2n+1} \mathcal{L}_{2p+2} \left(   \pi_{k}^{2},f\right),
\label{developpement commutateur moyal factorisation Ln ordre 1_2}
\end{align}
where the second term on the right-hand side reads as
\begin{align}
&\sum_{n=0}^{\infty} \hbar^{2n+2} \mathcal{L}_{2n+2}  \left(  \pi_{i}, \partial_{i} f\right)
=
 \frac{1}{\hbar}\sum_{n=0}^{\infty} \left(\frac{i \hbar}{2}\right)^{2n+3} \frac{\epsilon_{jlr} }{\left(2n+3\right)^{2} (2n+2)!} \left(\partial ^{2n+1}_{x_{i1} ... x_{i_{2n+1}}} B_{r} \right)    {}^L \! \partial_{\pi_{j}} {}^R \! \partial_{\pi_{l}} \nonumber \\
 &~~~
 \sum_{p=1}^{2n+1}
 \begin{pmatrix}
 2n+3 \\ p
 \end{pmatrix}
g(n,p)
~ {}^L \! \partial_{\pi_{i1}} \cdots   {}^L \! \partial_{\pi_{ip-1 }} {}^R \! \partial_{\pi_{ip}} \cdots   {}^R \! \partial_{\pi_{i_{2n+1} }}\left(  \pi_{i}, \partial_{i} f\right), \nonumber
\end{align}
with $g(n,p) =\left[\left(1-\left(-1\right)^{p}\right) \left(2n+3\right) - \left(1-\left(-1\right)^{2n+3}\right) p \right]$.\\
Only the term corresponding to $p=1$ gives a nonzero contribution
\begin{align}
\sum_{n=0}^{\infty} \hbar^{2n+2} \mathcal{L}_{2n+2}  \left(  \pi_{i}, \partial_{i} f\right)
&=
-\frac{i\hbar}{2}\sum_{n=0}^{\infty} \left(\frac{ \hbar}{2}\right)^{2n+1} \frac{(-1)^{n} (2n+2) }{(2n+3)!} \epsilon_{jlr}
\left(\partial ^{2n+1}_{i_{1} ... i_{2n+1}} B_{r} \right) \nonumber \\
&~~~ \partial ^{2n+1}_{\pi_{i_{1}} \cdots \pi_{i_{2n+1}}}  \partial_{\pi_{l}} \left(\partial_{j}f\right).
\label{result 1}
\end{align}
By similar developments, the third and the fourth terms of Eq. \eqref{developpement commutateur moyal factorisation Ln ordre 1_2} become
\begin{align}
\sum_{n=0}^{\infty} \hbar^{2n+1} \mathcal{L}_{2n+1}  \left(  \pi_{k}^{2}, f\right)
&=
 -\hbar \pi_{j}\epsilon_{jlr} \sum_{n=0}^{\infty} \left(\frac{ \hbar}{2}\right)^{2n} \frac{(-1)^{n} }{ (2n+1)!} \left(\partial ^{2n}_{i1 ... i_{2n}} B_{r} \right) \partial ^{2n}_{\pi_{i_{1}} \cdots \pi_{i_{2n}}}   \left(\partial_{\pi_{l}} f\right)
\label{result 2}
\end{align}
\begin{align}
&\sum_{n,p=0}^{\infty}  \hbar^{2n+1} \hbar^{2p+2}  \mathcal{L}_{2n+1} \mathcal{L}_{2p+2} \left(   \pi_{k}^{2},f\right)
=
\frac{i\hbar^{2}}{2} \epsilon_{ijk} \epsilon_{ilr}\sum_{n,p=0}^{\infty}  \left(\frac{ \hbar}{2}\right)^{2n+2p+1} \frac{ (-1)^{n}}{ (2n+1)!} \frac{(-1)^{p}(2p+2) }{(2p+3)!}\nonumber \\
&~~~
\left(\partial ^{2n}_{i1 ... i_{2n}} B_{r} \right)
\left(\partial ^{2p+1}_{j_1 ... j_{2p+1}} B_{k} \right)
\partial ^{2n}_{\pi_{i_{1}} \cdots \pi_{i_{2n}}}
\partial ^{2p+1}_{\pi_{j_{1}} \cdots \pi_{j_{2p+1}}}  \partial_{\pi_{j}}\left(   \partial_{\pi_{l}} f\right)
\label{result 3}
\end{align}
Injecting Eqs. \eqref{result 1}, \eqref{result 2} and  \eqref{result 3} into Eq. \eqref{developpement commutateur moyal factorisation Ln ordre 1_2}, one obtains :
\begin{align}
\left[\frac{\bm{\pi}^{2}} {2m}  , f_{0} \right]_{\star}
&=
-\frac{i\hbar}{m}\left[ \left(\bm{\pi} + \tilde{\bm{\Delta \pi}}  \right)\cdot \bm{\nabla}f\left(\bm{r},\bm{\pi},t\right)
-
e \left[\left(\bm{\pi} + \tilde{\bm{\Delta \pi}}  \right) \times \tilde{\bm{B}} \right]_{i} \partial_{\pi_{i}} f\left(\bm{r},\bm{\pi},t\right) \right],
\label{eq 1 f}
\end{align}
where we use the notation introduced by Serimaa et al. \cite{serimaa}:
\begin{align}
\bm{\Delta \widetilde{\pi}} &= -i\hbar e \partial_{\bm{\pi}} \times \left[ \int^{1/2}_{-1/2} d\tau \tau \bm{B} \left( \bm{r} + i\hbar \tau \partial_{\bm{\pi}} \right) \right],~~~~
\tilde{\bm{B}} =  \int^{1/2}_{-1/2} d\tau  \bm{B} \left( \bm{r} + i\hbar \tau \partial_{\bm{\pi}} \right).
\end{align}

\item  Term: $\left[  V, f_{0} \right]_{\star}$.

This term is the same as in the unmagnetized case. Indeed, the $\mathcal{L}_{n}$ operators do not act on $V$ since they contain at least one derivative in $\pi$. Thus, one obtains:
\begin{align}
&
\left[  V, f_{0} \right]_{\star}
=
\exp \left( i \hbar \mathcal{L} \right) \left[ \left( V,f_{0}(\bm{r},\bm{\pi}) \right) -  \left( f_{0}(\bm{r},\bm{\pi}) ,V\right) \right] \nonumber \\
&=
2i\sum_{n=0}^{\infty} \left( \frac{ \hbar}{2}\right)^{2n+1} \frac{(-1)^{n}}{(2n+1)!}  \left( {}^L \! \partial_{i} {}^R \! \partial_{\pi_{i}} - {}^R \! \partial_{j} {}^L \! \partial_{\pi_{j}}  \right)^{2n+1} \left( V,f_{0}(\bm{r},\bm{\pi}) \right),\nonumber \\
&= 2i \sum_{n=0}^{\infty}  \sum_{p=0}^{2n+1} \left( \frac{ \hbar}{2}\right)^{2n+1} \frac{(-1)^{n+p}}{(2n+1)!}
\begin{pmatrix}
2n+1\\p
\end{pmatrix}
 \left(  \partial_{i_{1} \cdots i_{2n+1-p}}^{2n+1-p} \partial_{\pi_{j_{1}} \cdots \pi_{j_{p}}}^{p} V \right)  \left( \partial_{\pi_{i_{1}} \cdots \pi_{i_{2n+1-p}}}^{2n+1-p} \partial_{j_{1} \cdots j_{p}}^{p} f_{0}\right).
 \label{eq pair}
\end{align}
Only the contribution corresponding to $p=0$ survives, because $V$ does not depend on $\bm{\pi}$. Thus:
\begin{align}
\left[  V, f_{0} \right]_{\star}
&=
-i\hbar e \bm{\widetilde{E}} \cdot \bm{\nabla_{\pi}} f_{0},
\label{eq 2 f}
\end{align}
where we have introduced the following quantity:
\begin{align}
\tilde{\bm{E}} = \int^{1/2}_{-1/2} d\tau \bm{E} \left( \bm{r} + i\hbar \tau \partial_{\bm{\pi}} \right).
\end{align}

\item  Term: $\epsilon_{ijk}  \left[  E_{i} \pi_{j} , f_{k} \right]_{\star}$.

We start by developing the following expression
\begin{align}
&
\exp \left(\i \hbar \mathcal{L} \right) \left(E_{i}\pi_{j},f_{k}\right)
=
\sum_{n=0}^{\infty} \left( \frac{i \hbar}{2}\right)^{n} \frac{1}{n!}  \left( {}^H \! \partial_{k} {}^f \! \partial_{\pi_{k}} - {}^f \! \partial_{l} {}^H \! \partial_{\pi_{l}}  \right)^{n}f_{k} H_{\mathcal{W}} ,\nonumber \\
&=
\sum_{n=0}^{\infty}  \sum_{p=0}^{n}  \left(\frac{i\hbar}{2}\right)^{n} \frac{(-1)^{p}}{n!}
\begin{pmatrix}
n\\p
\end{pmatrix}
 \left( \partial_{i_{1} \cdots i_{n-p}}^{n-p} \partial_{\pi_{j_{1}} \cdots \pi_{j_{p}}}^{p}   \pi_{j} E_{i}\right) \left( \partial_{\pi_{i_{1}} \cdots \pi_{i_{n-p}}}^{n-p} \partial_{j_{1} \cdots j_{p}}^{p} f_{k}\right).
\end{align}
Only the terms with $p=0$ or $p=1$ survive, so that
\begin{align}
\exp \left(\i \hbar \mathcal{L} \right) \left(E_{i}\pi_{j},f_{k}\right)
&=
\pi_{j} \sum_{n=0}^{\infty}   \left(\frac{i\hbar}{2}\right)^{n} \frac{1}{n!}
\left(  \partial_{\pi_{j_{1}} \cdots \pi_{j_{n}}}^{n}   f_{k}\right) \left(  \partial_{j_{1} \cdots j_{n}}^{n} E_{i}\right) \nonumber \\
&~
-\sum_{n=1}^{\infty}   \left(\frac{i\hbar}{2}\right)^{n} \frac{1}{(n-1)!}
\left( \partial_{\pi_{j_{1}} \cdots \pi_{j_{n-1}}}^{n-1}  \partial_{j} f_{k}\right) \left( \partial_{j_{1} \cdots j_{n-1}}^{n-1} E_{i}\right).
\end{align}
Then the Moyal product between $E_{i} \pi_{j}$ and $f$ can be written as:
\begin{align}
 \left(E_{i}\pi_{j}\right)\star f_{k}
&= \prod_{p=1}^{\infty}\exp \left( ie \hbar^{p} \mathcal{L}_{p} \right)\exp \left(\i \hbar \mathcal{L} \right) \left(E_{i}\pi_{j},f_{k}\right)
= \left[ 1+ie \sum_{p=1}^{\infty} \hbar^{p} \mathcal{L}_{p}\right] \exp \left(\i \hbar \mathcal{L} \right)  \left(E_{i}\pi_{j},f_{k}\right).
\end{align}
The last equality holds because the operator $\mathcal{L}_{p}$ acts one a phase-space function that is at most linear in $\bm{\pi}$.
Then we should evaluate the following quantity:
\begin{align}
&~ ie \sum_{p=1}^{\infty} \hbar^{p} \mathcal{L}_{p} \left[\exp \left(\i \hbar \mathcal{L} \right) \left(E_{i}\pi_{j},f_{k}\right)\right]
=
ie \sum_{n=0}^{\infty} \sum_{p=1}^{\infty} \hbar^{p}    \left(\frac{i\hbar}{2}\right)^{n} \frac{1}{n!}
\mathcal{L}_{p} \left(\pi_{j}\left(  \partial{j_{1} \cdots j_{n}}^{n} E_{i}\right), \left(  \partial_{\pi_{j_{1}} \cdots \pi_{j_{n}}}^{n}   f_{k}\right)\right) \nonumber \\
&~
=
ie \sum_{n=0}^{\infty} \sum_{p=1}^{\infty} \hbar^{p}    \left(\frac{i\hbar}{2}\right)^{n} \frac{1}{n!}
\left(\frac{i}{2}\right)^{p+1} \frac{\epsilon_{slr}}{(p+1)^{2}p!}\left(  \partial_{i_{1} \cdots i_{p-1}}^{p-1} B_{r}\right) \left(  \partial_{j_{1} \cdots j_{n}}^{n} E_{i}\right)
\sum_{m=1}^{p}
\begin{pmatrix}
p+1\\m
\end{pmatrix}
g(p,m)\nonumber \\
&~~~~
\left(  \partial_{\pi_{i_{1}} \cdots \pi_{i_{m-1}}}^{m-1} \partial_{\pi_{s}} \pi_{j}\right)
\left(  \partial_{\pi_{i_{m}} \cdots \pi_{i_{p-1}}}^{p-m}  \partial_{\pi_{j_{1}} \cdots \pi_{j_{n}}}^{n}  \partial_{\pi_{l}} f_{k}\right), \nonumber \\
&=
\frac{ie}{\hbar} \epsilon_{jlr} \sum_{n=0}^{\infty} \sum_{p=1}^{\infty}     \left(\frac{i\hbar}{2}\right)^{n} \frac{1}{n!}
\left(\frac{i\hbar}{2}\right)^{p+1} \frac{1}{(p+1)!}\left(  \partial_{i_{1} \cdots i_{p-1}}^{p-1} B_{r}\right) \left(  \partial_{j_{1} \cdots j_{n}}^{n} E_{i}\right)\nonumber \\
&~~~~
g(p,1)
\left(  \partial_{\pi_{i_{1}} \cdots \pi_{i_{p-1}}}^{p-1}  \partial_{\pi_{j_{1}} \cdots \pi_{j_{n}}}^{n}  \partial_{\pi_{l}} f_{k}\right), \nonumber
\end{align}
with $g(p,1) = 2p$ if $p$ is even and $g(p,1)=2(p+1)$ if $p$ is odd. Then one obtains:
\begin{align}
&
\left(E_{i}\pi_{j}\right)\star f_{k}
=
\sum_{n=0}^{\infty}  \frac{\pi_{j}}{n!}  \left(\frac{i\hbar}{2}\right)^{n}
\left(  \partial_{\pi_{j_{1}} \cdots \pi_{j_{n}}}^{n}   f_{k}\right) \left(  \partial_{j_{1} \cdots j_{n}}^{n} E_{i}\right) \nonumber \\
&~
-\sum_{n=1}^{\infty}   \left(\frac{i\hbar}{2}\right)^{n} \frac{1}{(n-1)!}
\left( \partial_{\pi_{j_{1}} \cdots \pi_{j_{n-1}}}^{n-1}  \partial_{j} f_{k}\right) \left( \partial_{j_{1} \cdots j_{n-1}}^{n-1} E_{i}\right)\nonumber \\
&~
+
\frac{ie}{\hbar} \epsilon_{jlr} \sum_{n=0}^{\infty} \sum_{p=1}^{\infty}     \left(\frac{i\hbar}{2}\right)^{n} \frac{1}{n!}
\left(\frac{i\hbar}{2}\right)^{p+1} \frac{1}{(p+1)!}\left(  \partial_{i_{1} \cdots i_{p-1}}^{p-1} B_{r}\right) \left(  \partial_{j_{1} \cdots j_{n}}^{n} E_{i}\right)
g(p,1) \nonumber \\
&~
\left(  \partial_{\pi_{i_{1}} \cdots \pi_{i_{p-1}}}^{p-1}  \partial_{\pi_{j_{1}} \cdots \pi_{j_{n}}}^{n}  \partial_{\pi_{l}} f_{k}\right),
\label{ei pij fk}
\end{align}
Using the symmetry identities \eqref{symmetric properties of l ln}, one obtains:
\begin{align}
&~
f_{k}\star \left(E_{i}\pi_{j}\right)
=
\sum_{n=0}^{\infty}   \frac{\pi_{j}}{n!}  \left(\frac{-i\hbar}{2}\right)^{n}
\left(  \partial_{\pi_{j_{1}} \cdots \pi_{j_{n}}}^{n}   f_{k}\right) \left(  \partial_{j_{1} \cdots j_{n}}^{n} E_{i}\right)
-\sum_{n=1}^{\infty}   \left(\frac{-i\hbar}{2}\right)^{n} \frac{1}{(n-1)!}\nonumber \\
&~
\left( \partial_{\pi_{j_{1}} \cdots \pi_{j_{n-1}}}^{n-1}  \partial_{j} f_{k}\right) \left( \partial_{j_{1} \cdots j_{n-1}}^{n-1} E_{i}\right)
+
\frac{ie}{\hbar} \epsilon_{jlr} \sum_{n=0}^{\infty} \sum_{p=1}^{\infty}     \left(\frac{i\hbar}{2}\right)^{n} \frac{(-1)^{n}}{n!}
\left(\frac{i\hbar}{2}\right)^{p+1} \frac{(-1)^{p}}{(p+1)!}\nonumber \\
&~
\left(  \partial_{i_{1} \cdots i_{p-1}}^{p-1} B_{r}\right) \left(  \partial_{j_{1} \cdots j_{n}}^{n} E_{i}\right)
g(p,1)
\left(  \partial_{\pi_{i_{1}} \cdots \pi_{i_{p-1}}}^{p-1}  \partial_{\pi_{j_{1}} \cdots \pi_{j_{n}}}^{n}  \partial_{\pi_{l}} f\right),
\label{fk ei pij}
\end{align}
Using Eqs. \eqref{ei pij fk} and \eqref{fk ei pij}, we finally obtain:
\begin{align}
&
 \frac{\epsilon_{ijk}}{i\hbar} \left[E_{i}\pi_{j},f_{k}\right]_{\star}
=
-\frac{1}{2}\left[ \left(\bm{E}_{+} + \bm{E}_{-} \right) \times \bm{\nabla} \right] \cdot \bm{f}
-
 \bm{\nabla} \left[\bm{\pi}  \times \widetilde{\bm{E}} \right]_{k} \cdot \bm{\nabla}_{\pi}  f_{k}
 \nonumber \\
&~
+e
\left[\widetilde{\bm{E}}  \times \left[ \widetilde{\bm{B}}  \times \bm{\nabla}_{\pi}\right] \right]\cdot \bm{f}
-\frac{1}{i\hbar}
\left[\bm{\Delta \widetilde{\pi}}  \times   \left(\bm{E}_{+} - \bm{E}_{-} \right) \right]  \cdot \bm{f}.
\label{eq 3 f}
\end{align}
The index $\pm$ means that the associated quantity is evaluated at a shifted position $ \bm{r} \pm i\hbar \partial_{\bm{\pi}} /2$.

\item Term: $ i \epsilon_{lri} \epsilon_{ijk} \left\{  E_{l} \pi_{r} , f_{j} \right\}_{\star}$

Using equation \eqref{ei pij fk} and \eqref{fk ei pij}, one directly obtains:
\begin{align}
&
 i \epsilon_{lri} \epsilon_{ijk} \left\{  E_{l} \pi_{r} , f_{j} \right\}_{\star}
=
-i \left\{ \left[ \left(\bm{\pi} + \bm{\Delta \widetilde{\pi}} \right)  \times    \left(\bm{E}_{+} + \bm{E}_{-} \right) \right] \times \bm{f} \right\}_{k} \nonumber \\
&~
 +
\frac{\hbar}{2} \left\{ \left[   \left(\bm{E}_{+} - \bm{E}_{-} \right) \times \left(\bm{\nabla} - e  \widetilde{\bm{B}} \times \bm{\nabla_{\pi}} \right)  \right] \times \bm{f} \right\}_{k}.
\label{eq 4 f}
\end{align}

\item Term: $ i \epsilon_{ijk} \left\{ B_{i},f_{j} \right\}_{\star}$

The calculation is very similar to that of the second item, i.e. for $\left\{ V ,f_{0} \right\}_{\star}$. The only difference is that we keep the even terms in Eq. \eqref{eq pair} instead of the odd terms. Then one simply obtains :
\begin{align}
i \epsilon_{ijk} \left\{ B_{i},f_{j} \right\}_{\star}
&= i \epsilon_{ijk} \left( \bm{B}_{+}  + \bm{B}_{-} \right)_{i} f_{j}.
\label{eq 5 f}
\end{align}

\end{enumerate}

Using the Eqs. \eqref{eq 1 f}, \eqref{eq 2 f}, \eqref{eq 3 f}, \eqref{eq 4 f}, \eqref{eq 5 f}, as well as the Eqs. \eqref{eq moyal f0}, \eqref{eq moyal f}, we finally obtain the Wigner equation for an electron interacting with an electromagnetic field, including the Zeeman interaction and the spin-orbit coupling:
\begin{align}
&
\frac{\partial f_{0}}{\partial t}
+
\frac{1}{m}\left(\bm{\pi} + \bm{\Delta \widetilde{\pi}}  \right)\cdot \bm{\nabla}f_{0}
-
\frac{e}{m} \left[ m\widetilde{\bm{E}} + \left(\bm{\pi} + \bm{\Delta \widetilde{\pi}}  \right) \times \widetilde{\bm{B}} \right]_{i} \partial_{\pi_{i}} f_{0}  \nonumber \\
&~
- \mu_{b} \bm{\nabla}\left( \widetilde{\bm{B}} -\frac{1}{2mc^{2}} \bm{\pi}  \times \widetilde{\bm{E}} \right)_{i}  \cdot \bm{\nabla_{\pi}} f_{i}
+
\frac{\mu_{b}}{4mc^{2}}\left[\left(\bm{E}_{+} + \bm{E}_{-}  \right)    \times \bm{\nabla} \right]  \cdot \bm{f} \nonumber \\
&~
-
\frac{\mu_{b}e}{2mc^{2}}
\left[\widetilde{\bm{E}}  \times \left[ \widetilde{\bm{B}}  \times \bm{\nabla}_{\pi}\right] \right]\cdot \bm{f}
-
\frac{\mu_{b}}{2mc^{2}}\frac{i}{\hbar}
\left[\bm{\Delta \widetilde{\pi}}  \times   \left(\bm{E}_{+} - \bm{E}_{-}  \right) \right]  \cdot \bm{f}
= 0, \label{wigner equation f0 condensee avec spin avec B ref}
\end{align}
\begin{align}
&
\frac{\partial f_{k}}{\partial t}
+
\frac{1}{m}\left(\bm{\pi} + \bm{\Delta \widetilde{\pi}}  \right)\cdot \bm{\nabla}f_{k}
-
\frac{e}{m} \left[ m\widetilde{\bm{E}} + \left(\bm{\pi} + \bm{\Delta \widetilde{\pi}}  \right) \times \widetilde{\bm{B}} \right]_{i} \partial_{\pi_{i}} f_{k}  \nonumber \\
&~
- \mu_{b} \bm{\nabla}\left( \widetilde{\bm{B}} -\frac{1}{2mc^{2}} \bm{\pi}  \times \widetilde{\bm{E}} \right)_{k}  \cdot \bm{\nabla_{\pi}} f_{0}
+
\frac{\mu_{b}}{4mc^{2}}\left[\left(\bm{E}_{+} + \bm{E}_{-}  \right)  \times \bm{\nabla} \right]_{k} f_{0} \nonumber \\
&~
-
\frac{\mu_{b}e}{2mc^{2}}
\left[\widetilde{\bm{E}}  \times \left[ \widetilde{\bm{B}}  \times \bm{\nabla}_{\pi}\right] \right]_{k} f_{0}
-
\frac{\mu_{b}}{2mc^{2}}\frac{i}{\hbar}
\left[\bm{\Delta \widetilde{\pi}}  \times   \left(\bm{E}_{+} - \bm{E}_{-}  \right)  \right]_{k} f_{0} \nonumber \\
&~
-\frac{e}{2m} \left[ \left(\bm{B}_{+} + \bm{B}_{-}  - \frac{1}{2mc^{2}} \left(\bm{\pi} + \bm{\Delta \widetilde{\pi}} \right)   \times \left(\bm{E}_{+} + \bm{E}_{-}  \right)\right) \times \bm{f}\right]_{k}\nonumber \\
&~
 +
 \frac{\mu_{b}}{2mc^{2}}\frac{i}{2}
 \left[ \left( \left( \bm{E}_{+} - \bm{E}_{-}  \right) \times \left( \bm{\nabla} -e  \widetilde{\bm{B}} \times \bm{\nabla_{\pi}} \right)  \right) \times \bm{f} \right]_{k} = 0.
\label{wigner equation f condensee avec spin avec B ref}
\end{align}